\documentclass[11pt]{article}
\usepackage{jheppub}

\usepackage{amsmath,amssymb, amsfonts}
\usepackage{graphicx}
\usepackage{comment}

\def\braket#1{\mathinner{\langle{#1}\rangle}}
\newcommand{\sbraket}[1]{\lbrack #1\rbrack}

\newcommand{\ii}{\mathrm{i}}

\newcommand{\al}{\alpha'}

\newcommand{\ha}{\frac{1}{2}}

\newcommand{\boxit}[1]{%
  \[\fbox{%
      \addtolength{\linewidth}{-2\fboxsep}%
      \addtolength{\linewidth}{-2\fboxrule}%
      \begin{minipage}{\linewidth}%
      #1%
      \end{minipage}%
    } \nonumber \]%
}

\title{Three particle superstring amplitudes with massive legs}
\author{Rutger H. Boels}
\affiliation{II. Institut f\"ur Theoretische Physik Universit\"at Hamburg\\ Luruper Chaussee 149, D- 22761 Hamburg, Germany}

\emailAdd{Rutger.Boels@desy.de}


\abstract{On-shell superspaces and associated spinor helicity techniques give an efficient formulation of the Ward identities of on-shell supersymmetry for scattering amplitudes and supply tools to construct their solutions. Based on these techniques in this paper the general solutions of the Ward identities are presented for three particle scattering amplitudes with one, two or three massive legs for simple supersymmetry in ten and eight dimensions. It is shown in examples how these solutions may be used to obtain concrete amplitudes for the closed (IIB) and open superstring in a flat background. Explicit results include all three point amplitudes with one massive leg whose functional form is shown to be dictated completely by super-Poincare symmetry. The resulting surprisingly simple series only involves massive superfields labelled by completely symmetric little group representations. The extension to more general explicit three and higher point amplitudes in string theory is initiated. In appendices the field content of the fundamental massive superfields of the open and closed superstring are listed in terms of the Dynkin labels of a variety of groups which may be of independent interest.}

\begin{document}
\maketitle

\section{Introduction}
The study of the scattering amplitudes it produces has taught us much of what is known about string theory today. The birth of the subject is after all widely taken to be the publication of Veneziano's amplitude \cite{Veneziano:1968yb}. This amplitude arose as an attempt to satisfy several constraints thought to hold for pion scattering amplitudes such as crossing symmetry and Regge behavior, without having to resort to off-shell (Lagrangian) methods. Precisely this ``analytic S-matrix'' philosophy has resurfaced in main-stream theoretical high energy physics after Witten's pioneering work \cite{Witten:2003nn}. In this new incarnation the main focus has been the calculation of scattering amplitudes of massless particles in (maximally supersymmetric) Yang-Mills theory and Einstein gravity in four dimensions. Considering the common analytic S-matrix roots a natural question is then if and how these new developments can be applied to string theory. 

This question was first raised in a series of papers by Stieberger and Taylor \cite{Stieberger:2007am}\cite{Stieberger:2007jv}\cite{Stieberger:2006te}\cite{Stieberger:2006bh}, where the focus was on calculating amplitudes with massless legs in the massless four dimensional spinor helicity formalism. A further step in this direction was taken in \cite{Boels:2008fc}. From the last reference evidence emerged that Britto-Cachazo-Feng-Witten (BCFW) on-shell recursion relations  \cite{Britto:2004ap}, \cite{Britto:2005fq} should arise very naturally in string theory. These relations allow one in principle to compute all scattering amplitudes from knowing the three point amplitudes for complex kinematics. Subsequently a proof of these relations in string theory appeared in two almost simultaneous articles \cite{Cheung:2010vn} \cite{Boels:2010bv} and have been explored somewhat for the Veneziano amplitude for instance in \cite{Fotopoulos:2010jz} and for four point superstring amplitudes in \cite{Feng:2011qc}. The string theory works mentioned in this paragraph are all at tree level and in a flat background. It should be noted though that the techniques in \cite{Cheung:2010vn} and \cite{Boels:2010bv} are phrased completely in CFT language and should generalize to much wider classes of backgrounds. Recently similar arguments have been investigated in the AdS/CFT context \cite{Raju:2011mp, Raju:2011ed}. In this paper the focus will be purely on tree level and on a flat background.

The main bottleneck to applying on-shell recursion in string theory as a practical tool is the fact that the relations involve sums over the full perturbative spectrum of the string theory. In particular this involves all massive Regge excitations. In principle it is known how to compute the needed three point `seed' amplitudes from the underlying CFT using the DDF operators, see  \cite{Schlotterer:2010kk} for instance for explicit expressions for the `highest spin' three point amplitudes. In practice, the resulting expressions are too complicated to be suited to use in an on-shell recursive calculation or indeed to yield any deeper insight in the structure of string perturbation theory. This result is somewhat counterintuitive, since the three point superstring amplitudes are in principle completely fixed by string theory's infinite dimensional symmetries which should lead to simple results. In order to bring out some of the underlying simplicity of the amplitudes in this article on-shell supersymmetry techniques are employed. In particular, on-shell superspaces are constructed in $10$ or $8$ dimensions for massive particles in the closed and open superstring respectively. Using these spaces explicit solutions to the supersymmetry Ward identities can be obtained. The case of supersymmetric amplitudes with only massless particles in these dimensions is considered in the companion article \cite{companion} to which the reader is also referred for more details  about the notation and spinor conventions. 

Three point amplitudes with generic matter content in the bosonic string have been obtained in \cite{Ademollo:1974kz} and more recently guessed in \cite{Fotopoulos:2010jz}. For the superstring similar work has appeared in  \cite{Schlotterer:2010kk} and \cite{Hornfeck:1987wt}. Higher point amplitudes with other than massless field content have not been discussed much in the literature, exceptions include \cite{Schlotterer:2010kk}, \cite{Liu:1987tb}, \cite{Xiao:2005yn}, \cite{Feng:2010yx}, and \cite{Feng:2011qc}. All expressions obtained in \cite{Feng:2010yx} have been unified in an explicitly supersymmetric form in \cite{Boels:2011zz} using an analysis of the four dimensional supersymmetry algebra for massive on-shell fields. In a seperate development, only recently an efficient method for obtaining the covariant massive string spectrum in terms of little group group superfield labels was obtained in \cite{Hanany:2010da}. The overall goal of which this article is a part is to compute all three point scattering amplitudes for the states in the spectrum of the string theory catalogued there in a reasonably simple form. If the so-called large extra dimensions scenario \cite{ArkaniHamed:1998rs} would be realized in string theory, then our results are directly relevant for collider experiments for the decay channels of massive string modes.

This article is structured as follows: in section \ref{sec:susyonshell} some relevant aspects of spinor helicity methods in eight and ten dimensions are introduced. New here is the explicit extension to general massive supermultiplets, fleshing out a general argument made in \cite{Boels:2009bv} in $D$ dimensions and in \cite{Boels:2011zz}  in four dimensions. In section \ref{app:derivfieldcontent} it is shown how the field content of the on-shell superfields can be calculated directly directly from the group structure of the relevant little groups. The on-shell superspaces constructed in section section \ref{sec:susyonshell} are applied in section \ref{sec:onemassiveleg} to derive forms for the scattering amplitudes with two massless and one massive leg in both open and closed superstrings up to normalization. It is shown explicitly how the derived expressions can be used to factorize the four massless particle amplitudes which also yields the normalization constants. Section \ref{sec:moregeneralampls} presents the solutions to the on-shell supersymmetry Ward identities for three point amplitudes with two or three massive legs. Illustrative example amplitudes based on these solutions are presented. Discussions and conclusions round off the main presentation. Appendices \ref{app:groupembeddingsD8} and \ref{app:groupembeddingsD10} display the field content of the massive fundamental multiplet in eight and ten dimensions in terms of a variety of groups.


\section{Massive on-shell supersymmetry in various dimensions}\label{sec:susyonshell}
Our main tool will be on-shell superspaces. As is well-known, representations of the on-shell supersymmetry algebra are technically much easier  to deal with than off-shell ones. It was realized first in \cite{Grisaru:1977px} in the massless four dimensional case that the representation theory of the linear on-shell supersymmetry algebra leads directly to Ward identities for scattering amplitudes, independently of any coupling constants. A general solution method for these Ward identities was introduced in \cite{Nair:1988bq} where the so-called MHV amplitude in four dimensional Yang-Mills amplitude was expressed in terms of variables on on-shell superspace. This also showed that the Ward identities are most naturally expressed in terms of four dimensional spinor helicity methods. These methods are a precise dictionary between vectors and Weyl spinors. A useful reference for the purposes of this article is the treatment of massive spinor helicity methods in four dimensions in \cite{Dittmaier:1998nn}.  

An extension of spinor helicity to six dimensions was presented in \cite{Cheung:2009dc}, followed by an extension of the susy Ward identities and superspaces in higher dimensions  in \cite{Boels:2009bv}. It was pointed out in \cite{Dennen:2009vk} that a particular superspace in six dimensions admits so-called supersymmetric delta function solutions to the Ward identities. In \cite{CaronHuot:2010rj} a discussion of the ten dimensional case is contained. The six dimensional superspace singled out by  \cite{Dennen:2009vk} was applied for instance in \cite{Bern:2010qa},  \cite{Brandhuber:2010mm}, \cite{Dennen:2010dh}. A companion paper \cite{companion} discusses the ten and eight dimensional spaces for massless particles at length. The reader is referred to that paper for a full explanation of the spinor notation employed below and details of the massless case. Here the focus is on the extension to massive particles. 

\subsection{Superspaces from massive D-dimensional spinor helicity}
The chiral supersymmetry algebra in $D$ dimensions reads
\begin{equation}
\{\overline{Q}_{A'},Q_{A}\} = k_{A'A}
\end{equation}
where the chiral (unprimed) and anti-chiral (primed) spinor indices run from $1$ to the dimension of chiral spinors in D dimensions, $\mathcal{D} = 2^{D/2-1}$. There is a choice of phases for solutions to the Dirac equation in $D$ dimensions such that a massless momentum $k^{\flat}_{A'A}$ can be expressed in terms of solutions to the massless Dirac equation as 
\begin{equation}
k^{\flat}_{A'A} = \lambda_{A', a'}  \lambda_{A}{}^{a'} 
\end{equation}
Here $a'$ denotes the anti-chiral spinor representation of the massless little group $SO(D-2)$. There is a special mass-less momentum $q$ connected to the choice of basis for the little group for which
\begin{equation}
q_{A'A} = \xi_{A'}{}^{a}  \xi_{A,a} 
\end{equation}
Here $a$ denotes the chiral spinor representation of the massless little group $SO(D-2)$. The $\lambda$ and $\xi$ spinors obey
\begin{equation}
[\xi_a \lambda^b] = n_{\lambda}  \, \delta_{a}{}^{b} \qquad  \langle \xi^{a'} \lambda_{b'} \rangle = \bar{n}_{\lambda} \,  \delta^{a'}{}_{b'} 
\end{equation}
as well as
\begin{equation}
[\lambda^{b' } \xi_{a'}] = m_{\lambda} \,  \delta^{a'}{}_{b'}  \qquad  \langle   \lambda_{a}  \xi^{b}\rangle = \bar{m}_{\lambda}   \, \delta_{a}{}^{b}
\end{equation}

Given a fixed massless momentum $q$, a massive momentum $k$ can always be decomposed into two massless momenta as 
\begin{equation}\label{eq:massivevecasspinors}
k_{A'A} = k^{\flat}_{A'A} + \frac{m^2}{2 q \cdot k} q_{A'A}
\end{equation}
Hence the supersymmetry algebra can  be written as
\begin{equation}\label{eq:massiveobshellsusy}
\{\overline{Q}_{A'},Q_{A}\} = \lambda_{A', a'}  \lambda_{A}{}^{a'}  + \frac{m^2}{2 q \cdot k} \xi_{A'}{}^{a}  \xi_{A,a} 
\end{equation}
Although it is possible in principle to choose different $q$'s to define little group indices and to define massless momenta, it is natural to choose both of these to be the same as will be done throughout this article. 

The main observation needed to go to an on-shell superspace is that fermionic multiplication and differentiation obey the fundamental relation
\begin{equation}
\left\{\eta, \frac{\partial}{\partial \eta} \right\} = 1
\end{equation}
Therefore, 
\begin{equation}\label{eq:qopera}
Q_A =  \lambda_{A}{}^{a'}  \eta_{a'} + \frac{m}{n_{\lambda}}\xi_{A,a}  \eta^a   \qquad \overline{Q}_{A'} =  \lambda_{A', a'}   \frac{\partial}{\partial \eta_{a'}} +  \frac{\bar{m}}{\bar{n}_{\lambda}} \xi_{A'}{}^{a}  \frac{\partial}{\partial \eta^{a} }
\end{equation}
is a complex representation of the on-shell supersymmetry algebra for massive momenta. There are $\mathcal{D}$ fermionic variables $\eta_{a'}$ as well as  $\mathcal{D}$ fermionic variables $\eta^{a}$.  Note that the chiral and anti-chiral Weyl representations of the massless little group $SO(D-2)$ can be combined into the Dirac representation of the massless little group $SO(D-2)$ which lifts to the spinor representation of the massive little group $SO(D-1)$. 

Possible Majorana conditions on the Weyl spinors do not influence the analysis: $\eta$ and $\frac{\partial}{\partial \eta}$ are naturally conjugate variables under Hermitean conjugation, so the constructed supersymmetry generators in equation \eqref{eq:qopera} are naturally complex. One can obtain other representations by interchanging multiplication and differentiation for a choice of little group indices. This can be implemented by fermionic Fourier transform on, say, variable $\eta_1$
\begin{equation}\label{eq:fermfourier}
F(\{\bar{\eta}_1 \eta_2, \ldots \})=  \int d\eta \, e^{ \bar{\eta}^1 \eta_1} F(\{\eta_1, \eta_2, \ldots\})
\end{equation}
These fermionic Fourier transforms will clearly break manifest little group transformation properties, unless applied to all fermionic variables on a leg simultaneously.

To each leg one can now associate an on-shell superfield, 
\begin{equation}
\phi(\eta, k_{\mu}) = \phi(k_{\mu}) + \eta_{a'} \phi^{a'} + \eta^a \phi_a  + \ldots + \left( \eta \right)^{2 \mathcal{D}} \overline{\phi}(k_{\mu})
\end{equation}
In general $\phi(k)$ has to transform in some representation of the massive little group which can be taken to be irreducible without losing generality. This is sometimes referred to as a choice of ``Clifford vacuum''. In general the other components in the superfield transform as a tensor product of the choice of Clifford vacuum with anti-symmetrized tensor products of the $SO(D-1)$ spinor representation. The anti-symmetrized tensor products contain $k$ spinors at fermionic level $k$. At the first level for instance, the $SO(D-2)$ indices on the fields on the first fermionic level form a $SO(D-2)$ Dirac spinor,
\begin{equation}
\phi_{\tilde{A}} = \left( \begin{array}{c} \phi^a \\ \phi_{a'} \end{array} \right)
\end{equation}
which lifts neatly to the spinor representation of $SO(D-1)$.

The superfields can therefore be taken to be labelled by the choice of Clifford vacuum: by irreps  (irreducible representations) of the massive little group. If this irrep is the scalar, the resulting superfield is referred to as the fundamental superfield. The little group transformation properties of the fields in the supermultiplet can then be worked out. See the next section and the appendices for worked-out examples in eight and ten dimensions. Under the fermionic Fourier transform of equation \eqref{eq:fermfourier} the content of the multiplet is shuffled around, generically breaking explicit little group symmetries. 

Since the representation of the supersymmetry algebra constructed is inherently complex and chiral it will be the massive representation of $\mathcal{N}=(2,0)$ (type IIB) in ten dimensions, while it is a massive representation of $\mathcal{N}=1$ in eight. For representations of extended supersymmetry algebra in a certain dimension one simply extends the fermionic variables as required. In principle one could aim to represent less supersymmetric theories such as $\mathcal{N}=(1,0)$ in ten dimensions as constrained superfields, this however leads beyond the scope of this paper.

\subsection{Ward identities of on-shell supersymmetry}
To make use of the representation of the supersymmetry algebra written in equation \eqref{eq:qopera} every leg of an amplitude is promoted to a superfield: each leg has bosonic momentum coordinate $k^{\mu}_i$ and fermionic `momenta'  $\eta_{i}^a,   \eta_{i,a'} $. Component amplitudes can be isolated by fermionic integration. The total momentum and supersymmetry generators are obtained as
\begin{equation}
K = \sum_i k_i \qquad Q_A = \sum_i Q_{A,i} \qquad \overline{Q}_{A'} = \sum_i  \overline{Q}_{A',i}
\end{equation}
Scattering amplitudes are required to be invariant under translations,
\begin{equation}
K A = 0
\end{equation}
as well as supersymmetry transformations
\boxit{\begin{equation}\label{eq:susywardids}
Q A = 0 \qquad \overline{Q} A =0
\end{equation}}
The last equations express the Ward identities of on-shell supersymmetry in a particularly concise form. The translation constraint leads to an overall momentum conserving delta function,
\begin{equation}
A \sim \delta(K^{\mu})
\end{equation} 
because the momentum operator is for momentum eigenstates (a.k.a. plane waves) purely multiplicative.

With all legs in the representation \eqref{eq:qopera} it is easy to solve parts of the second constraint as well as $Q$ in this representation is also purely a multiplication operator. Hence the amplitude must be proportional to the product of all these generators,
\begin{equation}
A \sim Q_1 Q_2 \ldots Q_{16} 
\end{equation}
where the subscript simply indicates spinor component. This can be phrased more covariantly in terms of the fermionic super-momentum conserving delta function
\begin{align}
\delta^{\mathcal{D}}(Q) 	& =  Q_1 Q_2 \ldots Q_{16}  \\
			& \equiv \frac{1}{\mathcal{D}!} \, \epsilon^{A_1 \ldots A_{\mathcal{D}}} \, Q_{A_1} \ldots Q_{A_{\mathcal{D}}} \label{eq:supermomcons}
\end{align}
In the second form it is obvious this function is invariant under Lorentz transformations. This function is invariant under $\overline{Q}$ up to momentum conservation by the supersymmetry algebra. Hence  
\begin{align}
\delta(K) \delta^{\mathcal{D}}(Q) 
\end{align}
is strictly invariant under both the $Q$ as well as the $\overline{Q}$ supersymmetry transformations. Therefore in this representation any amplitude can be written as
\begin{equation}\label{eq:genamplexp}
A = \delta(K) \delta^{\mathcal{D}}(Q) \tilde{A} \qquad \textrm{with} \qquad \overline{Q} \tilde{A} = 0
\end{equation} 
Hence solving the Ward identities for on-shell supersymmetry reduces to solving the equation on the right. The simplest solution to this is of course for $\tilde{A}$ to be independent of fermionic variables. 

One more use of on-shell superspace is to simplify the sum over all spins in a cut propagator as a fermionic integral,
\begin{multline}\label{eq:sumoverstates}
\sum_{spins}  A(X_L, \{P, \textrm{spin} \}) A(\{- P,\textrm{spin} \}, X_R) \quad \leftrightarrow \\  \int d\eta_P^{\mathcal{D}} A(X_L, \{\lambda_{A', a'}, \lambda_{A}{}^{a'}, \eta_P \}) A(\{\lambda_{A', a'}, -\lambda_{A}{}^{a'}, \eta_P \}, X_R)\
\end{multline}
The fermionic measure is defined as
\begin{equation}
\int d\eta^{\mathcal{D}} \equiv \frac{1}{(\mathcal{D}!)^2}  \int d \eta_{a'_1} \ldots d\eta_{a'_{\mathcal{D}/2}} d \eta^{a_1} \ldots d\eta^{a_{\mathcal{D}/2}} \epsilon_{a_1 \ldots a_{\mathcal{D}/2}} \epsilon^{a'_1 \ldots a'_{\mathcal{D}/2}} 
\end{equation}
The integral in equation \eqref{eq:sumoverstates} is a solution to the supersymmetry Ward identities in equation \eqref{eq:susywardids} if the amplitudes $A_L$ and $A_R$ are. 

\subsection{$U(1)_R$ symmetry conservation and fermionic weight}
As emphasized above, the representation of the on-shell supersymmetry algebra constructed in this section is inherently complex and admits an action of the $U(1)_R$ symmetry, 
\begin{equation}
Q\rightarrow e^{\ii \alpha} Q \qquad \overline{Q} \rightarrow e^{- \ii \alpha} \overline{Q}
\end{equation} 
This can be implemented by rotating the fermionic coordinates. The superfields can be given a consistent charge $q$ under this phase rotation by assigning opposite charges to the component fields at every fermionic level in the multiplet. It will turn out that these $U(1)$ charges are physical: in the eight dimensional case they are the eigenvalues under rotations in the $2$ extra directions present in the critical superstring. In the IIB ten dimensional theory this is part of the classical $SL(2,R)$ duality symmetry. We will assign all fermionic variables charge $1$. With this assignment conservation of $U(1)_R$ implies the fermionic weight of a scattering amplitude is simply the sum of the charges of the superfields 
\begin{equation}\label{eq:weightvscharge}
\textrm{fermionic weight }\left[ A_n \right] = \sum q_i
\end{equation} 
The massless superfields for instance have charge $2$ in the eight dimensional case and $4$ in the ten-dimensional case. The former follows from the identification of the scalar top-state with the extra-dimensional part of the ten dimensional gluon field. The latter follows from the $U(1)_R$ charges of the massless fields listed in the literature. 

Again, other representations of the massive superspace can be obtained by fermionic Fourier transform. In general this will lead to an expression for the scattering amplitudes which does not have fixed fermionic weight. An exception to this is the total fermionic Fourier transform which flips the entire multiplet.

An important constraint follows immediately from the above: only amplitudes with superfields whose R-charges sum to at least $\mathcal{D}$ are non-zero. Using the complete fermionic Fourier transform, it is also seen that there is a  maximal fermionic weight which depends on the number of massive and massless fields on the amplitude, 
\begin{equation}\label{eq:minmaxweight}
\mathcal{D} \leq \textrm{weight} \leq (\# \textrm{massless}) \frac{\mathcal{D}}{2} + (\# \textrm{massive}) \mathcal{D} -  \mathcal{D}
\end{equation}
This is especially restrictive in the three particle case. The case with three massless legs is exceptional as in this case (and in this case only) a weight $\frac{3}{4} \mathcal{D}$ delta function exists, see \cite{companion} for further details. 

\subsection{Explicit wave functions for amplitude computations}
Before continuing the main story about supersymmetry, it is useful to list some results for explicit wave functions. These can be used to compute explicit amplitudes from a field or string theory through standard methods. Wave functions for massless fields have been given in \cite{companion}. Here only the extension to massive fields will be listed. 

\subsubsection*{Massive vector polarizations}
 Consider a frame in $D$ dimensional space-time spanned by vectors $q,\bar{q}, n_i $ for $i=1,\ldots, (D-2)$, such that 
\begin{equation}
q \cdot \bar{q} = n_i \cdot n_i = 1 \qquad \textrm{no sum on i}
\end{equation}
are the only non-vanishing inner products. Then for a given on-shell massive momentum $k$
\begin{equation}
\tilde{n}_i(k) = n_i - \frac{q (n_i \cdot k)}{q \cdot k} 
\end{equation}
are a set of polarization vectors orthogonal to $k$ and $q$, while 
\begin{equation}
\tilde{n}^0_{\mu} = \frac{1}{m} \left(k_{\mu} - \frac{m^2}{q \cdot k} q_{\mu} \right)
\end{equation}
is the longitudinal polarization. It is clear that there is an $SO(8)$ which shuffles around the $n_i$ vectors (i.e. one can take the $i$ index to transform in the vector of $SO(8)$, while there are $SO(9)$ transformations involving the $n_i$ and $q$. The set $\tilde{n}_i$ for $i=0,\ldots, (D-2)$ is an orthonormal basis for the vector-space orthogonal to $k^{\mu}$, 
\begin{equation}
\sum_{i,j} \tilde{n}^i_{\mu} \tilde{n}^j_{\mu} \delta_{ij}  = \eta_{\mu\nu} - \frac{k_{\mu} k_{\nu}}{m^2} 
\end{equation}
by completeness. 

\subsubsection*{Massive Dirac spinor polarizations}
As expanded upon in \cite{companion},  $\lambda_{A', a'}$ and  $ \xi_{A'}{}^{a}$ form a basis for the primed spinors in $D$ dimensions, while  $\lambda^{A,a}$ and $\xi^{A}{}_{a'}$ form a basis of the unprimed spinors. A general Dirac spinor can be written in the chiral representation in this basis,
\begin{equation}
\psi = \left(\begin{array}{c} \psi^A \\  \psi_{A'} \end{array} \right) =  \left(\begin{array}{c}   d_a \lambda^{A,a} + d^{a'} \xi^{A}{}_{a'} \\  c^{a'} \lambda_{A',a'} + c_a \xi_{A'}{}^{a}\end{array} \right) 
\end{equation}
Set the mass matrix $m \!\!\! \slash$ to describe a complex chiral mass as
\begin{equation}
m \!\!\! \slash \equiv  m_1 \mathbb{I}_{\mathcal{D}} + \ii \,m_2\, \gamma_{D+1} = \left( \begin{array}{cc} m \,\mathbb{I}_{\mathcal{D}/2} & 0 \\ 0 & \bar{m}\,  \mathbb{I}_{\mathcal{D}/2} \end{array} \right)
\end{equation}
With the decomposition written in equation \eqref{eq:massivevecasspinors} it is easy to compute general solutions to the massive Dirac equation
\begin{equation}
(k \!\!\! \slash - m \!\!\! \slash ) \psi = 0 
\end{equation}
as
\begin{equation}
\psi_{a'} =    \left(\begin{array}{c} m \frac{\xi^{A}{}_{a'}}{m_{\lambda}} \\ \lambda_{A', a'} \end{array} \right) 
\end{equation}
and 
\begin{equation}
\psi^{a} =    \left(\begin{array}{c} \lambda^{A,a} \\ \bar{m} \frac{ \xi_{A'}{}^{a} }{\bar{m}_{\lambda}}\end{array} \right) 
\end{equation}
such that using equation \eqref{eq:massivevecasspinors} the following completeness relations holds
\begin{equation}\label{eq:masspolscomplet}
\psi^a \bar{\psi}_a + \psi_{a'} \bar{\psi}^{a'} = (k \!\!\! \slash + \bar{m}  \!\!\! \slash )
\end{equation}
where the bar indicates conjugation by the charge conjugation matrix. Note the neat interplay between little group spinor indices which conspire to make the Weyl spinors transform as little group spinors. The two chiral Weyl spinors of the $SO(D-2)$ massless little group again combine into the Dirac spinor of $SO(D-1)$. The massive vector and spinor polarizations form the basis of all polarization vectors. These are obtained by tensor products. This allows one to calculate scattering amplitudes in string and field theory with explicit little group transformation properties using textbook methods. 

In passing we note that the completeness relation for massive polarization spinors in equation \eqref{eq:masspolscomplet} can be used to construct natural BPS representations of the supersymmetry algebra through
\begin{equation}
Q = \psi^a \eta_a + \psi_{a'} \eta^{a'} \qquad \bar{Q} =  \bar{\psi}_a \frac{\partial}{\partial \eta_a }+ \bar{\psi}^{a'} \frac{\partial }{\partial \eta^{a'}}
\end{equation}
where the complex mass term gets reinterpreted as the matrix of central charges.

The natural minimal algebra which can be represented as such always entails twice as many super-symmetries as the massless case. Examples of this for the purposes of this article are for instance $\mathcal{N}=2$ in $D=8$ and $\mathcal{N}=(2,2)$ in $D=10$. Since dotted and un-dotted little group spinor indices in eight dimensions are equivalent, it can be seen that the ten dimensional superspace for massless multiplets is in fact equivalent to the just constructed BPS on-shell superspace in eight dimensions.


\section{Field content of massive 8 and 10 dimensional massive superfields}\label{app:derivfieldcontent}
The fields in the supersymmetric multiplet transform under the appropriate little group $SO(7)$ or $SO(9)$ in eight and ten dimensions respectively. It is instructive to derive the field content of the fundamental multiplets directly from group theory. This will be done in this section. As is well known irreducible representations of the simple Lie algebras under study can be uniquely specified by the Dynkin labels of a highest weight state. Let us therefore outline how to obtain these Dynkin labels for the superfields of the previous section. Below the focus in details will be on the ten dimensional case as the eight dimensional case is a simple restriction of this.

\subsection{Group theory strategy}
The little group transformation properties can be specified by the eigenvalues under rotations in the plane orthogonal to the massive on-shell momentum. This result also follows of course by going to the rest-frame of a single particle. These eigenvalues form the weight vectors of the little group. As outlined above, the superfields which will be studied in this article consist of anti-symmetrized tensor products of the little group spinor representation, so let us study this first. 

The weight labels of the little group spinor representation of the massless little group in ten dimensions, $SO(8)$, can be normalized to read for instance in the chiral case
\begin{equation}
\begin{array}{ccc}
& \left(\ha,\ha,\ha,\ha\right),\quad \left(-\ha,-\ha,\ha,\ha\right), \quad \left(-\ha,\ha,-\ha,\ha\right), \quad \left(-\ha,\ha,\ha,-\ha\right) & \\ 
& \left(\ha,-\ha,-\ha,\ha\right),\quad \left(\ha,-\ha,\ha,-\ha\right), \quad \left(\ha,\ha,-\ha,-\ha\right), \quad \left(-\ha,-\ha,-\ha,-\ha\right) & 
\end{array}
\end{equation}
and in the anti-chiral case
\begin{equation}
\begin{array}{ccc}
& \left(-\ha,\ha,\ha,\ha\right),\quad \left(\ha,-\ha,\ha,\ha\right), \quad \left(\ha,\ha,-\ha,\ha\right), \quad \left(\ha,\ha,\ha,-\ha\right) & \\ 
& \left(\ha,-\ha,-\ha,-\ha\right),\quad \left(-\ha,\ha,-\ha,-\ha\right), \quad \left(-\ha,-\ha,\ha,-\ha\right), \quad \left(-\ha,-\ha,-\ha,\ha\right) & 
\end{array}
\end{equation}
The sum of the chiral and anti-chiral weight labels are the labels of the spinor representation of $SO(9)$ (or the Dirac spinor representation of $SO(8)$). For the top-state of the so-called fundamental multiplet one picks the scalar which has weight labels
\begin{equation}
\vec{\lambda}_{\textrm{top}} = (0,0,0,0) \qquad \textrm{fundamental rep}
\end{equation}
in the ten dimensional case for instance. Alternatively, one can pick all states of a different representation as a top-state. The field content of the super-multiplet is then a tensor product of the top-state representation with the field content of the fundamental multiplet. Hence we can restrict attention first to this multiplet. 

The weights at some fermionic level follow from those of the one above by simply adding the weight labels of the chiral (or anti-chiral) little group spinor representation. Hence the weight labels of the anti-symmetrized tensor products can be calculated: the independent ones can be found by constructing ordered subsets of the weight labels of the $SO(9)$ spinor representation. At a specified fermionic level $n$ there are $\left(16 \atop n \right)$ or $\left(8 \atop n \right)$ different ways of doing this in the $10$ or $8$ dimensional case respectively. Using a symbolic manipulation program such as Mathematica one can generate at each fermionic level a list of all weight labels. 

The next step is to convert this list of weight labels to Dynkin labels, see for instance \cite{greenschwarzwitten} and the very brief review below. Within the set of Dynkin labels the highest weight states now have to be identified. These have positive Dynkin labels, so all other states may be dropped at this stage. States with positive Dynkin labels are called dominant weights. The set of Dynkin labels of the highest weight states is a subset of these: some of the dominant weights belong to states which occur inside another representation. To weed out the highest weight states, a list can be made of all dominant weights appearing inside a given highest weight state representation using computer algebra \cite{lie}.

The question now reduces to finding which combination of highest weight states and their dominant weights reproduces the found multiplicities. For massless supersymmetry this can be done by hand. In the massive case this has been done for the calculations in this article by sorting the Dynkin labels by the multiplicity with which they appear and using a spreadsheet program to keep track of the multiplicities. This step could be automated in principle. A useful consistency check is to calculate the sum of all dimensions of all the obtained highest weight state representations at a certain fermionic level and compare to the appropriate binomial coefficient.  The general procedure just outlined will be illustrated in an example below. 

\subsection*{Weight labels to Dynkin labels}
The simple roots of the algebra $D4$ ($SO(8)$) can be expressed in terms of the weight vectors $\vec{\lambda}$ as
\begin{equation}
\begin{array}{cc}
\alpha_1 & = (1,-1,0,0)\\
\alpha_2 & = (0,1,-1,0)\end{array} \qquad 
\begin{array}{cc}
\alpha_3 & = (0,0,1,-1)\\
\alpha_4 & = (0,0,1,1)
\end{array}
\end{equation}
The maximal set of commuting Chevalley basis generators follows from this as 
\begin{equation}
h_i = 2 \frac{\alpha_i}{(\alpha_i, \alpha_i)} R_i
\end{equation}
This defines the following map from weight labels $\vec{\lambda}$ to Dynkin labels $\vec{h}$, 
\begin{equation}
\begin{array}{cc}
h_1 & = (\lambda_1,-\lambda_2,0,0)\\
h_2 & = (0,\lambda_2,-\lambda_3,0)\end{array} \qquad 
\begin{array}{cc}
h_3 & = (0,0,\lambda_3, -\lambda_4)\\
h_4 & = (0,0,\lambda_3, \lambda_4)
\end{array}
\end{equation}

The simple roots of the group $B4$ ($SO(9)$) can be expressed in terms of the weight vectors $\vec{\lambda}$ as
\begin{equation}
\begin{array}{cc}
\alpha_1 & = (1,-1,0,0)\\
\alpha_2 & = (0,1,-1,0)\end{array} \qquad 
\begin{array}{cc}
\alpha_3 & = (0,0,1,-1)\\
\alpha_4 & = (0,0,0,1)
\end{array}
\end{equation}
This defines  the following map from weight labels $\vec{\lambda}$ to Dynkin labels $\vec{h}$, 
\begin{equation}
\begin{array}{ccc}
h_1 & = & \lambda_1 -\lambda_2\\
h_2 & = & \lambda_2 -\lambda_3\end{array} \qquad 
\begin{array}{ccc}
h_3 & = &\lambda_3 -\lambda_4\\
h_4 & = & 2  \lambda_4
\end{array}
\end{equation}
The different normalization of the last label is due to the non-simply laced algebra under study. Incidentally, by inversion of the above matrix one can transform $B4$ Dynkin labels into weight vectors as
\begin{equation}
\begin{array}{ccc}
\lambda_1 & = &  h_1 + h_2 + h_3 + \frac{1}{2} h_4 \\
\lambda_2 & = &  h_2 + h_3 + \frac{1}{2} h_4 \end{array} \qquad 
\begin{array}{ccc}
\lambda_3 & = &  h_3 + \frac{1}{2} h_4 \\
\lambda_4 & = & \frac{1}{2} h_4
\end{array}
\end{equation}
In particular, by the usual spin-statistics theorem, if and only if the fourth Dynkin label is an odd integer does the corresponding massive field in ten dimensions correspond to a fermion.

\subsection*{Example: massless superfield in ten dimensions}
After conversion of the weight labels to Dynkin labels of the $SO(8)$ little group and restricting to dominant weights the massless multiplet contains
\begin{equation}
\begin{array}{ccc}
0 & \quad &  (0,0,0,0)\\
1 & \quad &  (0,0,0,1) \\ 
2 & \quad &  4 \, (0,0,0,0) +  (0,1,0,0)\\
3 & \quad &  3 \, (0,0,0,1) +  (1,0,1,0) \\ 
4 & \quad &  6 \, (0,0,0,0) +  (0,0,2,0) + 2 \, (0,1,0,0) + (2,0,0,0)\\
5 & \quad &  3 \, (0,0,0,1) +  (1,0,1,0) \\ 
6 & \quad &  4 \, (0,0,0,0) +  (0,1,0,0)\\
7 & \quad &  (0,0,0,1) \\ 
8 & \quad &  (0,0,0,0)
\end{array}
\end{equation}
at the fermionic weight listed in the first column with the indicated multiplicities. Note that the fermionic weight $1$ level corresponds to the little group representation of the supersymmetry generator itself: the other levels are anti-symmetrized tensor products of this representation and in particular form themselves complete representations of the little group. The above list is symmetric under interchange of levels $i$ and $8-i$, as it should. 

From \cite{lie} one obtains the lists of dominant weights inside the highest weight state representations with the above Dynkin labels, sorted by the number of times the label appears in the lists, e.g. for the bosonic fields one obtains
\begin{equation}\label{eq:sortedlistofdominantweights}
\begin{array}{cccccc}
                & \quad 	& (2,0,0,0)& (0,0,2,0) 	& (0,1,0,0) & (0,0,0,0)	\\
(2,0,0,0) & \quad  	&       1      	&        0  		&        0 	&        0		\\
(0,0,2,0) & \quad      	&       0    	&        1   		&        0 	&        0   		\\
(0,1,0,0) & \quad       	&       1   	&        1     		&        1 	&        0  		\\
(0,0,0,0) & \quad    	&       3       &        3 		&        4 	&        1		 
\end{array}
\end{equation}
From this table and its fermionic partner one obtains at each fermionic weight in the massless fundamental multiplet complete highest weight state representations of $SO(8)$ as
\begin{equation}\label{eq:listofmasslessSO8reps}
\begin{array}{ccc}
0 & \quad &  (0,0,0,0)_{1}\\
1 & \quad &  (0,0,0,1)_{8} \\ 
2 & \quad &  (0,1,0,0)_{28}\\
3 & \quad &  (1,0,1,0)_{56} \\ 
4 & \quad &  (0,0,2,0)_{35} + (2,0,0,0)_{35}
\end{array}
\end{equation}
where the subscript indicates the dimension of the representation and each representation occurs once. Further levels repeat the pattern in reverse order. This is indeed the massless field content of type IIB supergravity. Interchanging chiral for anti-chiral spinors (i.e $(0,2)$ susy instead of $(2,0)$ susy) in the beginning of this calculation would yield a $(0,0,0,2)_{35}$ instead of $(0,0,2,0)_{35}$.

\subsection{Results for the massive case}
The calculation in the massive case is fundamentally the same as the above massless example, but due to the increased computational complexity of the calculation only results for the highest weight state representations will be listed here. 

\subsubsection*{Ten dimensions}
In the computation the equivalent of the matrix in equation \eqref {eq:sortedlistofdominantweights} contains $18$ entries on each side. The following result lists at each fermionic weight the fields in the supersymmetric multiplet in terms of representations of $SO(9)$, 
\begin{equation}\label{eq:dynkinlabelsso9}
\begin{array}{ccl}
0 & \quad &   (0,0,0,0)_{1}\\
1 & \quad &   (0,0,0,1)_{16} \\ 
2 & \quad &   (0,1,0,0)_{36} + (0,0,1,0)_{84}\\
3 & \quad &   (1,0,0,1)_{128} + (0,1,0,1)_{432}\\
4 & \quad &   (2,0,0,0)_{44} + (0,0,0,2)_{126} + (1,1,0,0)_{231} + (0,2,0,0)_{495} + (1,0,0,2)_{924}\\
5 & \quad &   (1,0,0,1)_{128} + (0,1,0,1)_{432} + (2,0,0,1)_{576} + (0,0,0,3)_{672}+ (1,1,0,1)_{2560} \\
6 & \quad &   (0,1,0,0)_{36} + (0,0,1,0)_{84}   + (1,1,0,0)_{231} + (1,0,1,0)_{594} + (1,0,0,2)_{924}  \\ & & + (2,1,0,0)_{910} + (2,0,1,0)_{2457} + (0,1,0,2)_{2772} \\
7 & \quad &   (0,0,0,1)_{16} + (1,0,0,1)_{128} + (0,1,0,1)_{432} + (2,0,0,1)_{576} \\ & & + (0,0,1,1)_{768} + (3,0,0,1)_{1920}+ (1,1,0,1)_{2560} + (1,0,1,1)_{5040}\\
8 & \quad &   (0,0,0,0)_{1} + (1,0,0,0)_{9} + (0,0,1,0)_{84} + (2,0,0,0)_{44} + (0,0,0,2)_{126} \\ & & + (1,0,1,0)_{594} + (0,2,0,0)_{495}   + (1,0,0,2)_{924} + (3,0,0,0)_{156} +  (0,1,1,0)_{1650} \\ & & + (2,0,1,0)_{2457}  + (2,0,0,2)_{3900}  + (0,0,2,0)_{1980} + (4,0,0,0)_{450}
\end{array}
\end{equation}
At a fixed fermionic weight each representation appears exactly once. The states appearing at fermionic weights $9$ through $16$ are the same as those at $7$ through $0$ respectively. A quick sanity check shows that the total number of states at level $n$ is equal to the binomial coefficient $\left(16 \atop n \right)$ as it should. By dimensional numerology, the total field content of the multiplet can be guessed to be given as the tensor square
\begin{equation}
\sim \left((2,0,0,0)_{44} + (0,0,1,0)_{84} + (1,0,0,1)_{128} \right)^{2}
\end{equation}
as this is the only combination of single (non-trivial) representations of $SO(9)$ whose dimensions sum to $256$ and can be split into bosonic and fermionic states evenly. It can of course also be verified directly. This square expresses the KLT relation  \cite{Kawai:1985xq} for free fields (see also below). In appendix \ref{app:groupembeddingsD10} embeddings of this field content into $SO(8)$, $SO(10)$, $SO(16)$, $SO(32)$ and $SO(3)\otimes SO(6)$ are given.

\subsubsection*{Eight dimensions}
The field content of the massive fundamental multiplet in eight dimensions can be calculated in terms of the $SO(7)$ ($B(3)$) little group to be:
\begin{equation}\label{eq:dynkinlabelsso7}
\begin{array}{ccl}
0 & \quad &   (0,0,0)_{1}\\
1 & \quad &   (0,0,1)_{8} \\
2 & \quad &   (0,1,0)_{21} + (1,0,0)_{7} \\
3 & \quad &   (1,0,1)_{48} + (0,0,1)_{8} \\
4 & \quad &   (2,0,0)_{27} + (0,0,2)_{35} + (1,0,0)_{7} + (0,0,0)_{1} \\ 
\end{array}
\end{equation}
where the unlisted fermionic levels repeat the previous ones in reverse order. In appendix \ref{app:groupembeddingsD8} embeddings of this field content into $SO(6)$, $SO(8)$, $SO(16)$  and $SO(3)\otimes SO(7)$ are given.

Again, the field content of the most general superfields in eight or ten dimensions is a tensor product of a representation of the appropriate little group with the above derived fundamental multiplets. Since the motivation of this technology is critical string theory all interesting multiplets arising in the eight dimensional case will be a dimensional reduction from ten dimensions: the superfields can be labelled by an $SO(9)$ representation. 

\subsection{KLT relations in eight dimensions on on-shell superspace}
The Kawai-Lewellen-Tye relations \cite{Kawai:1985xq} relate closed string scattering amplitudes to open string scattering amplitudes at tree level. For free fields this reduces to the standard observation that for instance the graviton polarization tensor can be written as a product of the polarization vectors of a massless vector boson. In the latter form this relation can be made manifestly supersymmetric in terms of the superfields introduced in this section.

For this one considers the dimensional reduction of the ten dimensional superfield to eight dimensions. This gives an $\mathcal{N}=2$ supersymmetric superfield in eight dimensions. To write this concisely one can decompose the ten dimensional superfield variables for each leg (indicated by variables with a tilde) in terms of the eight dimensional superfields variables as
\begin{equation}
\eta^{\tilde{a}} = \left(\begin{array}{c}  \eta^{a} \\ \eta_{a'} \end{array} \right) \qquad \eta_{\tilde{a}'} = \left(\begin{array}{c}  \bar{\eta}_{a} \\ \bar{\eta}^{a'} \end{array} \right)
\end{equation}
Here $\tilde{a}$ and ${\tilde{a}'}$ run from $1$ to $8$ while $a$ and $a'$ run from $1$ to $4$. The KLT relations in the free case simply state that the closed string superfield can be written in terms of open string superfields
\begin{equation}\label{eq:KLTfreesuperfield}
\Phi(\eta^{\tilde{a}}, \eta_{\tilde{a}'}) = \phi^{\textrm{D=8}}(\eta^{a} , \eta_{a'} ) \,\, \bar{\phi}^{\textrm{D=8}}(\bar{\eta}_{a} , \bar{\eta}^{a'} ) \qquad {\textrm{D=10 $\rightarrow$ 8}}
\end{equation}
on the product of a chiral and an anti-chiral superspace. These two superspaces are related by fermionic Fourier transform, 
\begin{equation}
\phi^{\textrm{D=8}}(\eta^{a} , \eta_{a'} )  = \int (d\eta_2)^8 e^{\eta^a \bar{\eta}_{a} + \eta_{a'} \bar{\eta}^{a'}} \bar{\phi}^{\textrm{D=8}}(\bar{\eta}_{a} , \bar{\eta}^{a'} ) 
\end{equation}
It can be verified explicitly (see equation \eqref{eq:KLTforfreefieldfieldcontent}) that the $SO(8)$ field content of the closed string superfield as given in subsection \ref{subsec:so8closed} can be written as the tensor product of the two open string superfields. The $SO(8)$ field content of the two open string superfields is given in equations \eqref{eq:so8contentopenstrchir} and \eqref{eq:so8contentopenstrantichir} respectively. Note that without the above total fermionic Fourier transform in eight dimensions this particular check does not work out: this can already be seen at fermionic level $1$ in the multiplet.


\section{Three point amplitudes with one massive particle}\label{sec:onemassiveleg}
In this section three point amplitudes with one massive leg are studied based on the on-shell superspace constructed above. An ingredient in this are the factorization channels of the four point massless superstring amplitudes constructed in \cite{companion}. These read
\begin{equation}
A^{D=8, \textrm{YM}}(G_1, G_2, G_3, G_4) =  \frac{\delta^8(K) \delta^8(Q)}{s \, t} \left[\frac{\Gamma\left(\al s + 1\right) \Gamma\left(\al t + 1\right) }{\Gamma\left(1 - \al u \right)} \right]
\end{equation}
in the eight dimensional case and
\begin{equation}
A_{4}^{D=10}(G_1, G_2, G_3, G_4)    =  \frac{\delta^{10}(K) \delta^{16}(Q)}{s \, t \, u}\left[\frac{\Gamma\left(\al s + 1\right) \Gamma\left(\al t + 1\right) \Gamma\left(\al u + 1\right) }{\Gamma\left(1-\alpha' s \right) \Gamma\left(1-\alpha' t  \right) \Gamma\left(1- \al u\right)} \right]
\end{equation}
in the ten dimensional case. Some of the component amplitudes contained in these forms describe open or closed superstring scattering of gluons or gravitons respectively. 

\subsection*{Fermionic weight from the amplitudes with four massless particles}
From equation \eqref{eq:minmaxweight} the scattering amplitudes with one massive leg must have fermionic weight $\mathcal{D}$. This can also be seen from factorizing the massless four point superstring amplitudes given above on a kinematic pole which involves a massive particle.  By tree level unitarity the result should be two three point amplitudes summed over the spectrum of the exchanged massive state. Schematically in say the s-channel one obtains in 10 dimensions,
\begin{equation}
\lim_{s \rightarrow -\frac{n}{\al}} A_4 = \sum_{I \in \textrm{superfields}}\int d\eta^{16} A_3(X_L, \{P, \eta\}^I) \frac{1}{s + \frac{n}{\al}} A_3(X_R, \{-P, \eta\}^I)
\end{equation}
where the integral contains the sum over states in \eqref{eq:sumoverstates}, $P$ is the sum of momenta on one side of the cut propagator and the sum over superfields ranges over all superfields which can appear at this mass-level. Generic superfields are labelled by the Dynkin labels of highest weight states. The sum over superfields includes the sum over all states in those highest weight state representations which couple through the indicated three point amplitudes. 

Since the four point amplitude $A_4$ has fermionic weight $16$, both three point amplitudes must have fermionic weight $16$. Hence by two arguments the three particle amplitude with one massive leg appearing in this factorization channel in the superstring must be proportional to the three point fermionic delta function,
\begin{equation}
A_3(1,2,3^I) =  C^I \delta^{16}(Q_1 + Q_2 + Q_3)
\end{equation}
This fact will also be proven generically in the next section. Here the capital $I$ stands for the little group Dynkin labels of the third, massive leg. Note that this assigns $U(1)$ R-charge $8$ to the massive ten dimensional superfield through equation \eqref{eq:weightvscharge}. It is particularly easy to integrate out all fermionic coordinates of the third leg,
\begin{equation}\label{eq:inttopstateonemass}
A(\phi_1, \phi_2, \Phi^I_3) = C^I \int d\eta_3^{16} \delta^{16}(Q_1 + Q_2 + Q_3) = C^I m^8
\end{equation}
Hence the three particle amplitude can be determined by computing an amplitude with two massless scalars and one massive state with little group quantum numbers $I$. Since the two scalars are bosonic, the massive superfield must be bosonic as well. The completely analogous computation in $8$ dimensions gives the result that two massless gluonic superfields couple to a massive bosonic superfield of R-charge $4$. The corresponding superamplitude is also proportional to the fermionic delta function.

\subsection{Open superstring in 8 dimensions}
Consider the case of open superstring in 8 dimensions. Before continuing it is useful to consider the interchange properties of the massless legs on a open superstring amplitude. For (any) two vertex operators on the boundary string worldsheet 
\begin{equation}
V(y_1) V(y_2) = V(y_2) V(y_1) \exp{\left(2 \pi \ii \al k_1 \cdot k_2 \theta(y_1 - y_2)\right)} 
\end{equation}
holds in a flat background, where $\theta(y)$ is the step function. For three particles in a color ordered amplitude this implies immediately that
\begin{equation}
A(1,2,3) = \exp(2 \pi \ii  k_1 \cdot k_2 ) A(2,1,3) = - (-1)^{\al \left(m_1^2 + m_2^2 + m_3^2 \right)}  A(2,1,3) 
\end{equation}
The extra minus comes from the stripped-off color trace. Since in string theory the masses are quantized in terms of $1/\al$ we obtain for the case where particles $1$ and $2$ are massless
\begin{equation}
A(1,2,3) = - (-1)^{n}  A(2,1,3) 
\end{equation}
where $n$ indicates the level of the massive state in the open superstring on the third leg. 

\subsubsection*{First excited level of the superstring}
It is well-known that in a flat background the open superstring has an infinite tower of massive excitations, where the masses are an integer multiple of $1/\alpha'$. See \cite{Hanany:2010da} for the most extensive list in terms of Dynkin labels of superfields the author is aware of. At the first excited level a scalar massive superfield appears. As argued above, the amplitude of this field and two massless superfields must be proportional to the weight eight fermionic delta function in eight dimensions, 
\begin{equation}
A(G,G,M) = C  \delta^{8}(Q_1 + Q_2 + Q_3)
\end{equation}
up to a bosonic function $C$ which is a little group scalar. This is symmetric under interchange of legs one and two as it should. From integrating out the fermionic coordinates on the massive leg, see equation \eqref{eq:inttopstateonemass}, one finds that the constant $C_M$ is a three point scalar scattering amplitude,
\begin{equation}
m^4 C_{M} = A(\phi^2, \phi^2, \Phi^{-4}) 
\end{equation}
where the superscripts indicate R-charge. The simplest guess for this amplitude is that it is a numerical constant. A more sophisticated argument is that the only functions of momenta allowed in here are the momentum invariants which in the three particle case are all constant by momentum conservation. Polarization vectors could only appear with all little group indices contracted which by completeness relations amounts to momentum and gauge vectors again. Since the result may not depend on a gauge choice, only the momentum may have a non-trivial influence on the amplitude which has however already been studied. 

On dimensional grounds this amplitude must be proportional to $\al \sqrt{\al}$. Hence
\begin{equation}\label{eq:eightDfirstRegge3pt}
A(G,G,M) = c \al \sqrt{\alpha'} \, \delta^{8}(Q_1 + Q_2 + Q_3)
\end{equation}
is the three point supersymmetric scattering amplitude with two massless legs and the first Regge excitation on the third leg, up to a numerical dimensionless constant $c$. Below this amplitude is compared to known results in the literature.

\subsubsection*{Second excited level of the superstring}
The second level superfield of the string contains a superfield with one $SO(9)$ vector index. This can be decomposed into $SO(7)$ representations as
\begin{equation}
(1,0,0,0) \rightarrow (1,0,0)^{0} + (0,0,0)^{\pm 2}
\end{equation}
where the superscripts indicate the $R$-charges. Hence the ten dimensional massive field content corresponds to a triplet of superfields,
\begin{equation}
M^{4,i}, M_+^{6}, M_-^{2}
\end{equation}
which are a $SO(7)$ little group vector indexed by $i$ and two $SO(7)$ scalars. The R-charges are fixed by the requirement that the spectrum be symmetric under R-charge conjugation, which follows from higher dimensional Lorentz symmetry. From equation \eqref{eq:minmaxweight} it follows that with one massive leg there is only one non-vanishing amplitude
\begin{equation}
A(G,G,M^{4,i}) = C^i \,  \delta^{8}(Q_1 + Q_2 + Q_3)
\end{equation}
In fact it is easy to translate the little group index on the third leg into a full Lorentz index using the massive polarization vectors on the third leg $e^{i}_{\mu}(3)$. Since the total amplitude is Lorentz invariant, this must be contracted with a Lorentz vector. This has to be some combination of $k_1$ and $k_2$. Since $k_3 \cdot e^{i}(3)= (k_1 + k_2) \cdot e^{i}(3) = 0$, this leads to 
\begin{equation}\label{eq:8dsuperspace3pt}
A(G,G,M^{i}) = c (\al)^2 \, (k_1 - k_2)_{\mu} e_3^{i,\mu} \, \delta^{8}(Q_1 + Q_2 + Q_3)
\end{equation}
as the only possible choice with $i$ the seven dimensional vector. This is anti-symmetric under interchange of legs $1$ and $2$ for this level of the string, as it should. The remaining numerical constant $c$ cannot be fixed from symmetry arguments alone, see below. Note that the above vector index can easily be extended to the nine dimensional vector index in the special little group basis where the extra two directions in the little group coincide with the extra two space-time directions from the lift to the ten-dimensional superstring. 

\subsubsection*{Third excited level of the superstring}
The third level superfield of the string contains two superfields labelled by the following $SO(9)$ Dynkin labels,
\begin{equation}
(2,0,0,0) \qquad (0,0,0,1)
\end{equation}
The first is the traceless symmetric representation while the second is the spinor representation. These can be decomposed into $SO(7)$ representations as
\begin{align}
(2,0,0,0) & \rightarrow (2,0,0)^{0} + (1,0,0)^{\pm 2} + (0,0,0)^{\pm 4} + (0,0,0)^{0} \\ 
(0,0,0,1) & \rightarrow (0,0,1)^{\pm 1} 
\end{align}
where the superscripts indicate the $R$-charges. Hence the ten dimensional field content corresponds to several bosonic superfields,
\begin{equation}
M^{4,ij}, \,M_+^{6,i},\, M_-^{2,i},\, M_+^{8},\, M_-^{0},\, M^{4}_0
\end{equation}
as well as a doublet of fermionic superfields,
\begin{equation}
M_+^{5,h},\, M_-^{3,h}
\end{equation}
with a $SO(7)$ spinor index $h$. With one massive leg there can be by R-symmetry conservation (equation \eqref{eq:minmaxweight}) only two different non-zero amplitudes,
\begin{equation}
A(M^{4,ij},G,G) \quad \textrm{and} \quad A( M^{4}_0,G,G)
\end{equation}
The first of these can from equation \eqref{eq:8dsuperspace3pt} immediately be guessed to be 
\begin{equation}\label{eq:cxasdc}
A(G,G,M^{ij}) \propto (\al)^2 \sqrt{\al}  \left[ \left((k_1 - k_2)_{\mu} e_3^{i,\mu} \right)   \left((k_1 - k_2)_{\mu} e_3^{j,\mu} \right) -  \frac{3}{\al} \frac{g^{ij}}{7} \right]\delta^{8}(Q_1 + Q_2 + Q_3)
\end{equation}
with $g^{ij}$ the seven dimensional little group metric, while the second is of the same form as the little group scalar at the first excited level of the superstring.  These two results must combine however to yield a quantity which transforms naturally under $SO(9)$. It is easy to guess that the normalizations of the amplitudes are such that 
\begin{equation}\label{eq:cxasdc2}
A(G,G,M^{ij}) =  c (\al)^2 \sqrt{\al} \left[ \left((k_1 - k_2)_{\mu} e_3^{i,\mu} \right)   \left((k_1 - k_2)_{\mu} e_3^{j,\mu} \right) -  \frac{3}{\al} \frac{g^{ij}}{9} \right]\delta^{8}(Q_1 + Q_2 + Q_3)
\end{equation}
where now the little group vector indices are nine-dimensional. That this is true follows from decomposing this form into seven dimensional irreps and evaluating the corresponding amplitudes in the special little group basis where the extra two directions in the little group coincide with the extra two space-time directions from the lift to the ten-dimensional superstring.

\subsubsection{Result for all excited levels}
The computations above lead to an immediate extension to amplitudes which involve a $j$-fold completely symmetric tensor products of the vector representation of $SO(9)$. 
\boxit{\begin{equation}\label{eq:threepointguess}
A(G,G,M^{i_1 \ldots i_j}) = \al (\sqrt{\al})^{j+1} \sqrt{\al} C(n,j) \left[ (k_1 - k_2)_{\mu} e_3^{i_1,\mu} \ldots (k_1 - k_2)_{\mu} e_3^{i_j,\mu} \right]\delta^{8}(Q_1 + Q_2 + Q_3)
\end{equation}}
Here $C(n,j)$ is a normalization constant. As a representation of $SO(9)$ the above tensor is not irreducible as it is not traceless in any of the indices. The dimension of this reducible $SO(9)$ representation can be expressed in terms of the binomial coefficient as
\begin{equation}
dim(j) = \left(9+j-1 \atop j \right)
\end{equation}
Note that the above form has the correct interchange properties for the massless legs $1$ and $2$ at even/odd $j$ for odd/even levels in the superstring. There are no other possibilities for a form which transforms only under the little group transformations of the massive leg. Hence this form is unique up to the normalization constant. 

Note that there is a special choice of `spin polarization axis' $q$ for the massive particle which makes \eqref{eq:threepointguess} even simpler: take $q=k_1$ or $q=k_2$. The only non-zero superamplitude left is the one where all polarization vectors in \eqref{eq:threepointguess} are longitudinal.

\subsubsection*{Comparison to known results}
The three point scattering amplitudes just considered have been calculated explicitly for the first Regge excitation in \cite{Feng:2010yx} and for the second excitation in \cite{Feng:2011qc} in the massive four dimensional spinor helicity language by worldsheet RNS methods. As will be shown now, these results are contained in equation \eqref{eq:threepointguess}. 

The results of \cite{Feng:2010yx} were translated into the massive four dimensional  $\mathcal{N}=4$ superfield language in \cite{Boels:2011zz}. For three points the amplitude was given there in two different superspaces:
\begin{equation}\label{eq:compfourdchiral}
A(G,G,M) = \frac{\sqrt{\al}}{\braket{12}^2} \,\,  \delta^{8}(Q^I_{\alpha}) \qquad D=4 
\end{equation}
in the so-called $\eta \bar{\iota}$ representation for all $4$ supersymmetries and
\begin{equation}\label{eq:compfourdantichiral}
A(G,G,M) =  \frac{\sqrt{\al}}{\sbraket{12}^2} \,\,  \delta^{8}(Q^I_{\dot{\alpha}}) \qquad D=4 
\end{equation}
in the so-called $\bar{\eta} \iota$ representation. Both these superspaces are (anti-)chiral and they are related by application of the fermionic Fourier transform on all superspace variables. This follows from applying
\begin{equation}
\int d\eta^I_1 d\eta^I_2 d\eta^I_3 d\bar{\iota}^I_3 e^{\bar{\eta}^I_i \eta^I_i + \bar{\iota}^I_3 \iota^I_3}\delta^{2}(Q^I_{\dot{\alpha}}) =  \frac{\braket{12}}{m}  \,\, \delta^{2}(\bar{Q}^I_{\alpha})
\end{equation}
written here for fixed $I$ four times and using $\braket{12}\sbraket{12} = m^2 =-\frac{1}{\al}$ by momentum conservation. To compare to the higher dimensional result above one should transform two of the four fermionic variables in either equation \eqref{eq:compfourdchiral} or \eqref{eq:compfourdantichiral}. The result is:
\begin{equation}
A(G,G,M) =  -  \al  \sqrt{\al}\,\, \delta^{4}(Q^I_{\alpha}) \delta^{4}(Q^J_{\dot{\alpha}}) \qquad D=4 
\end{equation}
where $I=1,2$ and $J=3,4$. It is easy to check that the dimensional reduction of the amplitude in equation \eqref{eq:eightDfirstRegge3pt} exactly reproduces this result. Hence in this particular case the proposed three point amplitude with one massive leg reproduces known results obtained by vertex operator methods in the open string. 

The results in \cite{Feng:2011qc} for the second excited level of the superstring can be summarized by the one-line superamplitude
\begin{equation}\label{eq:compfourdchiral2ndex}
A(G,G,M^i) = \frac{\sqrt{\al}}{\braket{12}^2} \,\, (k_1 - k_2)^{\mu} e^{3,i}_{\mu} \,\,  \delta^{8}(Q^I_{\alpha}) \qquad D=4 
\end{equation}
in the chiral ( $\eta \bar{\iota}$) superspace. This can be Fourier transformed to the non-chiral superspace
\begin{equation}\label{eq:compfourdnonchiral2ndex}
A(G,G,M^i) =  - \al \sqrt{\al}  \,\, (k_1 - k_2)^{\mu} e^{3,i}_{\mu} \,\,  \delta^{4}(Q^I_{\alpha}) \delta^{4}(Q^J_{\dot{\alpha}}) \qquad D=4 
\end{equation}
which is the dimensional reduction of equation \eqref{eq:8dsuperspace3pt}.

\subsubsection*{Relation to Dynkin labels}
As a representation of the little group the $3$ point amplitudes of equation \eqref{eq:threepointguess} transform as a symmetrized tensor product of the vector representation. It is instructive to realize that the Dynkin labels of the traceless symmetric $\lambda$ index tensor of $SO(9)$ are
\begin{equation}
(\lambda,0,0,0)
\end{equation}
One way to check this is to calculate the dimension of the traceless symmetric tensor representation in nine dimensions, which is
\begin{equation}\label{eq:dimtracsym}
\textrm{dim[traceless,symmetric, $\lambda$ indices ]} = \left(\begin{array}{c} 9+ \lambda - 1 \\ \lambda \end{array} \right) - \left(\begin{array}{c} 9+ \lambda - 3 \\  \lambda -2 \end{array}  \right)
\end{equation}
The first binomial coefficient is the dimension of the symmetric $l$ index tensor, while the second subtracts of the number of equations in the traceless-ness condition. This is to be compared to the dimension of the $(\lambda,0,0,0)$ representation of $SO(9)$ as calculated using the dimension formula which is in turn derived from Weyl's character formula,
\begin{equation}\label{eq:dimb4rep}
\textrm{dim}\left[(\lambda,0,0,0)\right] = \frac{(\lambda+1)(\lambda+2)(\lambda+3)(\lambda+4)(\lambda+5)(\lambda+6)(2\lambda+7)}{5040}
\end{equation}
The positive roots for the algebra $B(4)$ needed in the general dimension formula were obtained from \cite{lie}. Comparing equation \eqref{eq:dimtracsym} and equation \eqref{eq:dimb4rep} shows perfect agreement for all $\lambda$.

Note that this calculation shows that the three point scattering amplitudes of equation \eqref{eq:threepointguess} only involve a subset of the full spectrum of the string as listed in \cite{Hanany:2010da}. Some of this is simply because the corresponding superfields in eight dimensions would be fermionic (recognizable by their last odd integer entry in the Dynkin label) and cannot couple to two bosonic states in a three point amplitude. It would be interesting to understand this point better.

\subsection{Factorizing the open superstring amplitude with four massless gluons}
The normalization of the amplitudes in equation \eqref{eq:threepointguess} can be obtained by a study of the factorization properties of the massless four point open superstring amplitude in eight dimensions which by tree level unitarity reads 
\begin{equation}\label{eq:factvend8}
\lim_{s \rightarrow -\frac{n}{\al}} A_4^{D=8} = \sum_{\textrm{superfields}}\int d\eta^{16} A_3(X_L, \{P, \eta\}) \frac{1}{s+\frac{n}{\al}} A_3(X_R, \{-P, \eta\})
\end{equation}
Here it will be shown the class of amplitudes written in \eqref{eq:threepointguess} is indeed enough to describe all factorization channels. From the explicit expression for the massless amplitude from \cite{companion} one obtains
\begin{equation}\label{eq:kinpolefourpointopen}
\lim_{s \rightarrow -\frac{n}{\al}} A_4^{D=8} \rightarrow  \frac{(-1)^{n}}{\Gamma(n)} \al \left(  \frac{1}{s + \frac{n}{\al}} \right) \frac{\Gamma(\alpha' t)}{\Gamma(\alpha' t - n + 1)}\delta^8(Q)
\end{equation}
Comparing the last two equations shows that there is information on the massive spectrum hidden in the polynomial 
\begin{equation}\label{eq:factpoly}
\frac{(-1)^n}{\Gamma(n)} \al \frac{\Gamma(\alpha' t)}{\Gamma(\alpha' t - n + 1)}
\end{equation}
since this should be equal to the superfield sums in equation \eqref{eq:factvend8}. The integral over the fermionic variables in that equation can be performed once and for all,
\begin{multline}
\sum_I \int d\eta_P^{8}  A_3(1,2,P^I) A_3(-P^I,3,4) = \\ (\frac{n}{\al})^2 \, \sum_I C(1,2,P^I) C(-P^I,3,4) \delta^8(Q_1+Q_2+Q_3+Q_4)
\end{multline}
with some numerical constant $x$, leaving a contraction of little group indices indicated by $I$. Referring back to our guess in equation \eqref{eq:threepointguess} this is seen to involve multiple summations over little group indices of the type
\begin{equation}
\sum_{ij} e^{i}_{\mu}(P) e^{j}_{\nu}(P) \delta_{ij} = g_{\mu\nu} - \frac{P_{\mu} P_{\nu}}{P^2}
\end{equation}
which holds by the completeness relation for massive vector polarization vectors. Contracted with the relevant momentum factors from  \eqref{eq:threepointguess} this yields as a part of the residue calculation at the kinematic pole under study
\begin{equation}
\al \sum_{ij} (k_1 - k_2)^{\mu} e^{i}_{\mu}(P) e^{j}_{\nu}(P) \delta_{ij} (k_3 - k_4)^{\nu} =  2 \al t -  n
\end{equation}
Hence a completely symmetric tensor superfield of rank $j$ (which is \emph{not} an irrep) will contribute a polynomial in $t$ to the residue in the form of
\begin{equation}\label{eq:jfactorsfact}
\sim \left(2 \al t - n \right)^j
\end{equation}
up to the overall normalization constant of the superfield squared. Comparing with the known form of the polynomial \eqref{eq:factpoly} which appears in the residue at the kinematic pole of the massless four point amplitude from equation \eqref{eq:kinpolefourpointopen} then yields the field content of the massive states of the open superstring as well as the normalization of the three point functions up to some subtleties worked out below. This follows from the equation
\begin{equation}\label{eq:detfieldcontent}
x \frac{n^2}{\al^2} \al \sum_I C(1,2,P^I) C(-P^I,3,4) = \alpha' \frac{(-1)^n}{\Gamma(n)}  \frac{\Gamma(\alpha' t)}{\Gamma(\alpha' t - n + 1)}
\end{equation}
It will now be shown the three point amplitudes in equation \eqref{eq:threepointguess} are sufficient to satisfy \eqref{eq:detfieldcontent}. This can be rephrased as a question about polynomials. For even $n$ for instance, the factorization  
\begin{equation}
\frac{\Gamma(\alpha' t)}{\Gamma(\alpha' t - n + 1)} \stackrel{?}{=} \sum_{j=0}^{n/2} c^2_j  \left(2 \al t - n \right)^{2j} n^2
\end{equation}
should hold. Note that the correct interchange properties for the massless legs $1$ and $2$ at even/odd $j$ for odd/even levels in the superstring give a restriction on the polynomial. Therefore this is a non-trivial relation as there are half the number of coefficients $c^2_j$ as there are coefficients in the polynomial on the left hand side. Before continuing with the general analysis some examples will be presented first.

\subsubsection*{Level 1}
At this level the polynomial on the right hand side of equation \eqref{eq:detfieldcontent} is a constant. Hence there is only the fundamental superfield at the first massive level whose scattering superamplitude with the massless fields reads
\begin{equation}
A(G,G, M_1) =  \al \sqrt{\al} \ii \delta^{8}(Q_1 + Q_2 + Q_3) 
\end{equation}

\subsubsection*{Level 2}
At this level the polynomial on the right hand side of equation \eqref{eq:detfieldcontent} is linear. This equation can be satisfied simply using a vector superfield. Hence the second level of the open superstring contains a vector superfield  whose scattering superamplitude with the massless fields reads
\begin{equation}
A(G,G, M^i_2) =  \frac{1}{2} \frac{1}{\sqrt{2}} \al^2 (k_1 - k_2)_{\mu} e_3^{i,\mu} \delta^{8}(Q_1 + Q_2 + Q_3) 
\end{equation}

\subsubsection*{Level 3}
At this level the polynomial on the right hand side of equation \eqref{eq:detfieldcontent} is quadratic. Hence the second level of the superstring contains a up to a two index superfield: either a scalar superfield or the symmetric-traceless superfield in terms of irreps of $SO(9)$.  The first step is to factorize the polynomial on the right hand side of \eqref{eq:detfieldcontent} into $(2 \al t - 3)$ factors,
\begin{equation}
2^{(2)} (\al t - 2)(\al t -1) = (2 \al t - 3)^2 - (2 \al t - 3)^0
\end{equation}
The $(2 \al t -3)^j$ factors arise from superfields with $j$ indices as in equation \eqref{eq:jfactorsfact}. Hence at level $3$ there can be a two index symmetric field as well as a zero index symmetric field. However, the first of this is not an irrep of $SO(9)$. The scattering amplitude for the trace-less symmetric superfield (which is an irrep) reads:
\begin{multline}
A(G,G, M^{ij}_3) = \frac{\ii}{3} \frac{1}{\sqrt{\Gamma(3)}} \al \sqrt{\al} \\ \left[\al \left((k_1 - k_2)_{\mu} e_3^{i,\mu\phantom{\beta}}\! \!\right)  \left((k_1 - k_2)_{\mu} e_3^{j,\mu} \right) - \frac{3}{9} \delta^{ij} \right] \delta^{16}(Q_1 + Q_2 + Q_3) 
\end{multline}
A short calculation will show that this is indeed traceless in the $ij$ indices. Putting this amplitude into the factorization formula \eqref{eq:detfieldcontent} shows that the trace-less symmetric superfield is the only superfield at level 3 which couples to two bosonic massless fields. The other superfield which is known at this level is a fermionic spin $\frac{1}{2}$ superfield which does not couple to the two bosonic fields. 

\subsubsection*{Level 4}
The polynomial factorization into $\al t -2$ factors required reads
\begin{equation}
 (\al t -3)(\al t - 2)(\al t -1) = ( \al t - 2)^3 -  (\al t - 2)^1
\end{equation}
hence a combination of the three index symmetric and one index superfields. The trace-less symmetric three index field amplitude reads
\begin{multline}
A(G,G, M^{ijk}_3) =  \frac{ \al^2 }{2} \frac{1}{\sqrt{\Gamma(4)}}\left[ \al \left( (k_1 - k_2)_{\mu} e_3^{i,\mu\phantom{\beta}}\! \!\right)  \left( (k_1 - k_2)_{\nu} e_3^{j,\nu} \right) \left((k_1 - k_2)_{\rho} e_3^{k,\rho\phantom{\beta}}\!\! \right)  - \right. \\ \left. \frac{4}{11}\left( \delta^{ij}   \left(  (k_1 - k_2)_{\mu} e_3^{k,\mu} \right) + \textrm{cyclic}   \right) \right] \delta^{16}(Q_1 + Q_2 + Q_3) 
\end{multline}
From the factorization formula it is seen that at the fourth excited level there therefore must also be at least one $SO(9)$ vector representation superfield coupling to the massless modes, just as it did at the second level. Based on the basis of the arguments presented it is unfortunately not possible to measure the multiplicity of this vector superfield. 

\subsection*{All levels}
After these examples and some experimentation with Mathematica it is seen that there are always integer coefficients $\tilde{c}_j$ such that 
\begin{equation}
2^{(n-1)} \frac{\Gamma(\al t)}{\Gamma(\al t-n+1)} = \sum_{j=0}^{(n-1)/2} (2 \al t -n)^{2j} \tilde{c}_j 
\end{equation}
or
\begin{equation}
2^{(n-1)} \frac{\Gamma(\al t)}{\Gamma(\al t-n+1)} = \sum_{j=1}^{n/2} (2 \al t -n)^{2j-1} \tilde{c}_j
\end{equation}
for the $n$ odd or even cases respectively. The resulting coefficients can be identified with the central factorial numbers with a little help from \cite{njas}. From this knowledge one can work out a relation to their generating function in the $n$ odd case for instance as follows:
\begin{align}
2^{(n-1)} \frac{\Gamma(\al t)}{\Gamma(\al t-n+1)} & = \prod_{i=1}^{(n-1)/2} ((2 \al t - n) - i) ((2 \al t - n) +  i) \\
& = \prod_{i=1}^{(n-1)/2}((2 \al t - n)^2 - i^2) = \sum_{i=0}^{(n-1)/2} f_i (2 \al t - n)^{2i} \label{eq:defgenfac}
\end{align}
The last line is the (normalized) generating functional of the central factorials $f_i$. A similar relation follows in the odd case. This equation demonstrates that in the factorization of the massless superstring amplitude one can interpret the arising residues at the pole in terms of the three point couplings of equation \eqref{eq:threepointguess} only.

To summarize, at a fixed level $n$ one needs superamplitudes for the completely symmetric vector representations of the $SO(9)$ little group as given in equation \eqref{eq:threepointguess} with up to $n-1$ little group indices, i.e.
\begin{align}
n \, = \, \textrm{even} & \rightarrow j=1,3,\ldots, n-1 \\
n \, = \, \textrm{odd} & \rightarrow j=0,2,\ldots, n-1 
\end{align} 
The normalization constant $C(n,j)$ of the amplitudes in equation \eqref{eq:threepointguess} is given by in the $n$ odd case by
\begin{equation}
C(n,j) = \sqrt{\frac{f_{j/2}}{n \, 2^{n-1}\, \Gamma(n) \}}}
\end{equation}
where the coefficients $f_i$ are defined in equation \eqref{eq:defgenfac} and in the $n$ even case by
\begin{equation}
C(n,j) = \sqrt{\frac{f_{(j-1)/2}}{n\, 2^{n-1} \, \Gamma(n) }}
\end{equation}
in the $n$ odd case. 

It is interesting to compare the found three point couplings to the full field content of the superstring as obtained in for instance \cite{Hanany:2010da} to relatively high excitation levels. It should be clear from this comparison that only a very small subset of modes of the string couple to two massless modes: namely only those whose Dynkin labels read $(\lambda,0,0,0)$, with $\lambda$ smaller or equal to the level of the string mode. Furthermore, $\lambda$ is even or odd if the level of the string is odd or even respectively. 
 
\subsection{Closed IIB superstring in ten dimensions}
The extension to three particle amplitudes with one massive state of the closed superstring in $10$ dimensions follows from the above open string analysis by the KLT relations \cite{Kawai:1985xq}. For this one needs to realize that all essentially eight dimensional formulas in the above only involve the fermionic delta functions. The necessary generalization to 10 dimensions of this is rather straightforward, 
\boxit{\begin{multline}
A(G,G,M^{i_1 \ldots i_{2j}}) = \left(\sum_{j_1 + j_2 = j} C(n,j_1)C(n,j_2)\right) \left(\al\right)^{j+3} \\ \left[ (k_1 - k_2)_{\mu} e_3^{\alpha_1,\mu} \ldots (k_1 - k_2)_{\mu} e_3^{\alpha_j,\mu} \right]^{2 j} \delta^{16}(Q_1 + Q_2 + Q_3)
\end{multline}}
since this has the right dimensional reduction to eight dimensions. At level $n$ in the string the closed string sector contains fields with $j=0,2,\ldots, 2(n-1)$ completely symmetric $SO(9)$ vector indices. The sum over $j_1$ and $j_2$ ranges over the allowed values of $j$ on the open superstring amplitude in equation \eqref{eq:threepointguess}. Note that the resulting closed string amplitude is symmetric under interchange of the two massless legs, as it should. 

In principle the same analysis of the factorization channels of the closed superstring amplitude can be given as above in the open case. This however quickly reduces to two copies of the above by expressing the Virasoro-Shapiro factor in terms of two Veneziano factors through
\begin{multline}
\left[\frac{\Gamma\left(\al s + 1\right) \Gamma\left(\al t + 1\right) \Gamma\left(\al u + 1\right) }{\Gamma\left(1-\alpha' s \right) \Gamma\left(1-\alpha' t  \right) \Gamma\left(1- \al u\right)} \right] =\\ \frac{\textrm{sin}(\pi \alpha' t)}{\pi  \alpha' t}  \left[\frac{\Gamma\left(\al s + 1\right) \Gamma\left(\al t + 1\right) }{\Gamma\left(1- \al u \right)} \right] \left[\frac{\Gamma\left(\al t + 1\right) \Gamma\left(\al u + 1\right) }{\Gamma\left(1- \al s \right)} \right]
\end{multline}
which follows from Euler's reflection identity and the observation that the ten-dimensional four point fermionic delta function when reduced to eight dimensions explicitly factorizes.


\section{Three point amplitudes with two and three massive legs}\label{sec:moregeneralampls}
This section discusses the extension of the above results to more general three point amplitudes. It is useful to first consider the possible fermionic weights. In eight dimensions it is clear that fermionic superfields have odd integer R-charge while bosonic superfields have even integer R-charge. Through the KLT relation \eqref{eq:KLTfreesuperfield} and dimensional reduction this fact translates to the closed string. Since fermions should always appear pairwise on amplitudes, this implies that the fermionic weight of all superamplitudes have to be even integers. 

\begin{table}
\centering
\begin{tabular}{c|ccc}
\# massive particles & open & & closed \\
\hline
0 &  6  & &12 \\ 
1 &  8  & &16\\
2 &  8, 10, 12 & & 16, 18, 20, 22,24 \\
3 &  8, 10, 12, 14, 16 & & 16, 18, \ldots \ldots, 30, 32

\end{tabular}
\caption{Possible fermionic weights of three point superamplitudes in the open and closed string cases as a function of number of massive particles \label{tab:allfermweigopen}}
\end{table}
Moreover, from equation \eqref{eq:minmaxweight} the bounds on minimal and maximal fermionic weight are fairly stringent for three particles. The possibilities are listed in table \ref{tab:allfermweigopen} in the open and closed string cases. Note that the amplitudes are symmetric with respect to the `middle' fermionic weight: using complete Fourier transform once a certain amplitude is found, a conjugate amplitude follows. With two massive particles in the open string case for instance, scattering amplitudes with fermionic weight $12$ are related to those with weight $8$ by complete fermionic Fourier transform. 

\subsection{All solutions to the Ward identities for three particles}
First it will be shown that the on-shell supersymmetry Ward identities can be solved for three particle amplitudes in general. From equation \eqref{eq:genamplexp} such a solution can be constructed from a polynomial function $\tilde{A}$ for which
\begin{equation}\label{eq:defoftildeA}
\overline{Q}_{B'} \tilde{A} = 0 \qquad \textrm{and} \qquad \tilde{A} \sim \tilde{A} + f(Q)
\end{equation}
for any polynomial function $f$ of the supermomentum $Q_{B}$. The latter follows from the supersymmetric delta function. In the three particle case there are always three sets of $\mathcal{D}/2$ variables
\begin{equation}
\{\eta_{1,a'}, \eta_{2,a'}, \eta_{3,a'} \}
\end{equation}
while there are zero, one, two or three sets of variables $\eta^{a}$ depending on the number of massive legs. Generically, the freedom of shifting in equation \eqref{eq:defoftildeA} can be used to set to zero two out of three of the $\eta_{i,a'}$ variables. Then from 
\begin{equation}
\xi^{A',a} \overline{Q}_{A'} \tilde{A} = 0 
\end{equation}
it follows that the remaining $\eta_{a'}$ variable can also be set to zero, as long as all massive legs are based on the same spinors $\xi$. Hence the function $\tilde{A}$ can be taken to be a function of the $\eta^{a}$ variables of the massive legs only. In the one massive leg case this implies directly that there is no non-trivial fermionic function as the remaining constraint reads 
\begin{equation}
\xi_{A'}{}^a \frac{\partial}{\partial \eta^a} \tilde{A} = 0 \qquad  \textrm{one massive leg}
\end{equation}
which only leaves non-fermionic functions $\tilde{A}$. This reconfirms the analysis of this case in the previous section.

\subsubsection*{Two massive legs}
In the case of two massive legs the remaining equation after taking into account the fermionic delta function is
\begin{equation}
\xi_{A'}{}^a \left(\frac{\bar{m}_1}{\bar{n}_1} \frac{\partial}{\partial \eta_{1}^a} + \frac{\bar{m}_3}{\bar{n}_3} \frac{\partial}{\partial \eta_{3}^a} \right)\tilde{A} = 0 \qquad \textrm{two massive legs}
\end{equation}
where legs one and three are taken to be massive. Since by a simple coordinate transformation $\tilde{A}$ can be written as a polynomial function 
\begin{equation}
\tilde{A} \sim \tilde{A}(\frac{\bar{n}_1}{\bar{m}_1} \eta_{1}^{a} \pm \frac{\bar{n}_3}{\bar{m}_3}  \eta_{3}^{a} ) 
\end{equation}
the constraint is solved by
\begin{equation}\label{eq:solwardid2mass}
 \tilde{A} \sim \prod_{i=1}^{w}  (\frac{\bar{n}_1}{\bar{m}_1} \eta_{1}^{a} - \frac{\bar{n}_3}{\bar{m}_3}  \eta_{3}^{a} )
\end{equation}
up to a multiplicative bosonic function which can be a function of momenta and little group indices. Note that the minimal and maximal values of $w$ are $0$ and $4$ respectively. This solution transforms as a $w$-fold anti-symmetrized tensor product of the massless little group chiral spinor representation ($D(4)/D(3)$ in the closed and open string cases). In the ten dimensional case the two- and four-fold anti-symmetric tensor product of the massless little group representations can be read off from the field content in \eqref{eq:listofmasslessSO8reps}. In the eight dimensional case the equivalent two-fold anti-symmetric tensor product is related to the field content of the massless gluonic superfield. This also immediately gives the number of degrees of freedom in the solution to the Ward identities: for the solution in \eqref{eq:solwardid2mass} this is the binomial coefficient $\left(\mathcal{D}/2 \atop w \right)$ in general dimensions. Calculating this number of different amplitudes for the field content of the involved superfields fixes all the other superamplitudes which are related by supersymmetry.

Generically it is hard to see how with $w$ odd the resulting massless little group tensor can be used to describe the scattering of massive string states. This is an ``on-shell'' mirror of a conclusion reached before: the superamplitudes should have even fermionic weight. To use the above solution to the Ward identities to construct non-vanishing amplitudes some sense needs to be made of the loose massless little group indices. Since there is one field on the amplitude which is massless it is natural to use this to translate the massless little group indices into full Lorentz indices. These then have to be contracted into some Lorentz tensor to yield a nice invariant amplitude. 

\subsubsection*{Three massive legs}

The supersymmetric Ward identity in the case of three massive legs reduces to a differential equation of the type
\begin{equation}
\left( \frac{\partial}{\partial \tilde{\eta}_{1}^{a}} + \frac{\partial}{\partial \tilde{\eta}_{2}^a} + \frac{\partial}{\partial \tilde{\eta}_{3}^a} \right) \tilde{A}(\eta_1, \eta_2, \eta_3)
\end{equation}
where for convenience the variables $\eta_{i,a}$ have been rescaled to $\tilde{\eta}_{i}^a \equiv \frac{\bar{n}_i}{\bar{m}_i} \eta_{i}^a$. By a change of variables,
\begin{equation}
 \tilde{A}(\eta_1, \eta_2, \eta_3) \rightarrow  \tilde{A}(\tilde{\eta}_1 + \tilde{\eta}_2 + \tilde{\eta}_3, \tilde{\eta}_1 - \tilde{\eta}_2, \tilde{\eta}_2 -\tilde{\eta}_3)
\end{equation}
so that the general solution is a polynomial of two variables of fixed weight, say
\begin{equation}
\tilde{A}(\tilde{\eta}_1 - \tilde{\eta}_2, \tilde{\eta}_2 -\tilde{\eta}_3)
\end{equation}
Since in the open string case the interchange properties of external legs are important, it is in this case sometimes useful to express the amplitude as a function of dependent variables
\begin{equation}\label{eq:solwardid3mass}
\tilde{A}(\tilde{\eta}_1 - \tilde{\eta}_2, \tilde{\eta}_2 -\tilde{\eta}_3, \tilde{\eta}_3-\tilde{\eta}_1)
\end{equation}
As in the case of two massive legs in general the anti-symmetrized tensor products of the massless little group involved do not lift to representations of the massive little group.

The counting of the number of degrees of freedom is slightly more complicated in this case. If the function $\tilde{A}$ has fermionic weight $w$, then this number is 
\begin{equation}
\# \textrm{DOF} = \sum_{j=0}^{w} \left(\mathcal{D}/2 \atop j \right) \left(\mathcal{D}/2 \atop w-j \right)
\end{equation}
Special symmetries for special field content on the amplitudes may reduce this considerably. Calculating this number of different amplitudes for the field content of the involved superfields fixes all the other superamplitudes which are related by supersymmetry. A table of the total number of degrees of freedom in eight and ten dimensions is listed in table \ref{tab:degoffreedcount}. 

\begin{table}
\centering
\begin{tabular}{c|ccccc}
           	& 0 &  2 &  4 & 6 & 8 \\
\hline
D=8 		&  1 &  28 &  70 & 28 & 1  \\ 
D=10 	&  1 &  120 &  1820 & 8008 & 12870
\end{tabular}
\caption{ Degrees of freedom left in solutions of the (simple) supersymmetry Ward identities with three massive particles in ten and eight dimensions as a function of fermionic weight\label{tab:degoffreedcount}}
\end{table}

\subsection{Explicit examples in the open string}
This subsection displays examples of explicit scattering amplitudes based on the found solutions of the Ward identities with more than one massive leg. Since closed string amplitudes follow from open string ones by KLT, the focus will be on the latter in eight dimensions.

\subsubsection{Two massive superfields}
Consider first the scattering amplitude of two R-charge $4$ massive scalar superfields from the first excited level and one massless gluonic field.  From the reasoning above, this amplitude is proportional to 
\begin{equation}\label{eq:twomassscalaronegluon}
A(M_1 G_2 M_3) \sim L_{ab} (\tilde{\eta}^a_1 - \tilde{\eta}^a_3)(\tilde{\eta}^b_1 - \tilde{\eta}^b_3) \delta^8(Q)
\end{equation}
where the little group factor $L_{ab}$ has to be determined, for instance by calculating $6$ explicit amplitudes. In fact, the easiest thing to do is to integrate out all fermionic variables on, say, leg $3$ and two of them on leg $2$ as
\begin{equation}
A(\phi  g^{a'b'} \bar{\phi}) = \int d\eta_{2,a'} d\eta_{2,b'} d\eta_3^8 \,\, A(M_1 G_2 M_3)
\end{equation}
The resulting amplitude is simply of scalar-gluon-anti-scalar type, where the gluon can assume $6$ polarization states. Assuming minimal coupling, this amplitude is known. In fact, it has been worked out in eight dimensions for similar purposes in \cite{companion}. This gives 
\begin{equation}
A(M_1 G_2 M_3) = \frac{L_{ab}}{L_{ab} [\xi^a |\slash \!\!\! 1 \slash \!\!\! 3 |\xi^b]} (\tilde{\eta}^a_1 - \tilde{\eta}^a_3)(\tilde{\eta}^b_1 - \tilde{\eta}^b_3)  \delta^8(Q)
\end{equation}
as a natural result for the three point amplitude with the indicated field content, up to a numerical factor. The only constraint on $L$ is that it should not make the denominator vanish.

Based on the previously obtained results one can immediately guess broad generalizations of the above amplitudes. From \eqref{eq:twomassscalaronegluon} for instance, one can based on similarity to the one-massive state case in equation \eqref{eq:threepointguess} construct
\boxit{\begin{multline}\label{eq:threepointguess2mass}
A(M_1,G_2, M_3^{i_1 \ldots i_j}) \stackrel{?}{=} \al (\sqrt{\al})^{j+k+1} \sqrt{\al} C(n,j,k) \left[ (k_1 - k_2)_{\mu} e_3^{i_1,\mu} \ldots (k_1 - k_2)_{\mu} e_3^{i_j,\mu} \right] \\A(M_1 G_2 M_3)
\end{multline}}
as a candidate amplitude for the scattering of one massive completely symmetric superfield and one massive scalar superfield. Even more generally, for two completely symmetric superfields one can construct a similar ansatz. In this case the polarization vectors of the two massive legs can be contracted into momentum factors as above or into each other. This is easy to argue based on the structure of the CFT vertex operators which would lead to equation \eqref{eq:threepointguess}. This suggests that for two completely symmetric superfields with $j$ and $k$ indices respectively one would obtain 
\begin{equation}\label{eq:contractgensupfields}
\sim \left(\al (k_1 - k_2)_{\mu}  (k_3 - k_2)_{\nu} + \eta_{\mu\nu} \right)e_1^{i,\mu} e_3^{i,\nu}
\end{equation}
for the ``overlapping'' number of indices ($i<j\leq k$ or $i<k\leq j$). This has of course to be appropriately symmetrized. Along similar lines one can try to construct scattering amplitudes with two fermionic superfields and extensions thereof.

\subsubsection{Three scalar superfields with equal masses}
The final amplitude of this paper involves three scalar superfields with equal masses. These superfields have each have charge $4$ and the resulting superamplitude has to be proportional to the solution to the Ward identity given in equation \eqref{eq:solwardid3mass}. As calculated above, there are $70$ degrees of freedom in this amplitude in general. However, in this special case the amplitude has to be completely symmetric under interchange of any of the two legs. Moreover, the resulting combination has to be a little group scalar. It can be checked that the only little group scalars which can arise in the weight four polynomial in equation \eqref{eq:solwardid3mass} arise by contracting all little group indices with the completely anti-symmetric four index tensor $\epsilon_{a'b'c'd'}$. Hence the superamplitude reduces to 
\begin{multline}\label{eq:scalarthreefieldansatz}
\tilde{A} \sim  \left(a_0(\eta_1 - \eta_2)^4 + a_1 (\eta_1 - \eta_2)^3(\eta_2 - \eta_3) + \right. \\ \left. a_2 (\eta_1 - \eta_2)^2(\eta_2 - \eta_3)^2  + a_3 (\eta_1 - \eta_2)(\eta_2 - \eta_3)^3 + a_4 (\eta_2 - \eta_3)^4 \right)
\end{multline}
for five parameters $a_i$, which is as far as we can get with three general massive scalar superfields. Here the spinor indices have been suppressed. Complete permutation symmetry then fixes these parameters up to an overall constant which we take to be $a_0$ to
\begin{equation}
\{a_1, a_2, a_3, a_4\} \rightarrow \{2 a_0, 3 a_0, 2 a_0,  a_0\}
\end{equation}
which can be rewritten as 
\begin{equation}
\tilde{A} \sim  \left((\eta_1 - \eta_2)^4 + (\eta_2 - \eta_3)^4 + (\eta_3 - \eta_1)^4 \right)
\end{equation}
Hence this scattering amplitude is fixed up to a constant which can be calculated from the scattering amplitude
\begin{equation}
A(\phi \tilde{\phi} \bar{\phi})
\end{equation}
where the $\tilde{\phi}$ field is the sole scalar at fermionic level four in the massive supermultiplet, see equation \eqref{eq:dynkinlabelsso7}. Since this is an amplitude only involving scalars it must be proportional to a scalar function. Since there is nothing to compare to yet this will be left to future work. An admittedly ambitious guess of the extension to more generic completely symmetric superfields reads 
\boxit{\begin{multline}\label{eq:threepointguess3mass}
A(M,M, M^{i_1 \ldots i_j}) \stackrel{?}{\sim} \left[(k_1 - k_2)_{\mu} e_3^{i_1,\mu} \ldots (k_1 - k_2)_{\mu} e_3^{i_j,\mu}\right]\\
 \left[ (\eta_1 - \eta_2)^4 + (\eta_2 - \eta_3)^4 + (\eta_3 - \eta_1)^4 \right]\delta^8(Q)
\end{multline}}
More general three point amplitudes with completely symmetric superfields would be expected to involve the tensor contraction in equation \eqref{eq:contractgensupfields} in various constellations. Note that for three massive scalar superfields with two masses equal to each other it is natural to have the same ansatz as in equation \eqref{eq:scalarthreefieldansatz} with particles $2$ and $3$ of equal mass. The permutation constraint then fixes 
\begin{equation}
\{a_1, a_2, a_4\} \rightarrow \{\pm a_3, \pm a_2, \pm a_0\}
\end{equation}
depending on the level of the third massive field. Hence there are three constants left to fix in this case. 

\subsubsection*{On obtaining the correct normalizations}
To obtain the normalization of the three point scattering amplitudes described in this subsection one would like to have four point scattering amplitudes with one massive leg: the factorization channels of this can then be used just as was done above. Since it is known from explicit results these take the form of a homogeneous term in $\alpha'$ times a Veneziano factor, this can in principle be calculated using on-shell recursion with the results known above. For this one simply examines only the massless pole term from the recursion relation to obtain the pre-factor of the Veneziano factor, which has fermionic weight two on our superspace. 

Alternatively and perhaps ultimately more straightforwardly these normalizations may in principle be calculated directly from the CFT description of the superstring.


\section{Discussion and Conclusions}
In this article we have introduced and applied on-shell superspaces for massive multiplets in eight and ten dimensions. These superspaces arise from the representation theory of the on-shell supersymmetry algebra and are useful for formulating all-orders solutions to the SUSY Ward identities. We have pointed out that specific minimal solutions exist on these superspaces: these are the supersymmetric delta function which are invariant under all supersymmetry variations. These solutions are in the three point case enough to formulate explicit scattering amplitudes. As an example all three point amplitudes with one massive and two massless legs have been obtained. As a cross-check and to obtain the correct normalization the factorization of amplitudes with four massless fields on massive poles has been studied. 

The discussion above of more general three point couplings form the first step toward obtaining all three point amplitudes in the open and closed superstring. This is an obvious further research direction. It should be interesting to understand the selection rules exposed in the above three point amplitudes better: for instance by obtaining the $R$ charges of the closed string superfields in the classification of \cite{Hanany:2010da}. Moreover, it would be interesting to see the structure of higher point amplitudes with massive legs. This can be obtained through on-shell recursion relations at least in principle. For this one would use higher dimensional supersymmetrized BCFW relations, generalizing the four dimensional construction of \cite{Brandhuber:2008pf} and \cite{ArkaniHamed:2008gz}. See \cite{companion} for more details. Studying the systematics of on-shell recursion for, say, gluon or graviton scattering amplitudes also remains a goal. The simplifications found in the three point amplitudes above give hope this might be simpler than previously thought. In fact the similarities between different amplitudes suggests to try to reformulate, say, the completely symmetric fields in the full spectrum of the superstring into something like an ``on-shell string field''.

A natural question raised by the results in this article is how to write supersymmetric expressions for super-Yang-Mills amplitudes in ten dimensions: the minimal superfield in ten dimensions constructed here has too many fermionic degrees of freedom. Actually, a way around this is well-known in off-shell representations on the lightcone \cite{Brink:1983pf}: the Yang-Mills superfield needs to be constructed as a constrained on-shell superfield.
  
More theoretically, the results found in this paper should also be produced directly by any CFT description of the superstring. After all, the three point amplitudes are the ``structure constants'' of the CFT. It would for instance be interesting to see for instance how the RNS computations in \cite{Schlotterer:2010kk} or especially more space-time supersymmetric pure spinor computations compare to the results obtained here. Along similar lines, one could try to use the results of this article to constrain and determine the corresponding parts of the string theory effective action. The description of off-shell higher spin fields is however notoriously difficult.

\section*{Acknowledgements}
It is a pleasure to thank Donal O'Connell for collaboration on related work \cite{companion} for massless multiplets, Christoph Horst and and Oliver Schlotterer for discussions and Jan de Boer and Bernard de Wit for useful comments. This work was supported by the German Science Foundation (DFG) within the Collaborative Research Center 676 ``Particles, Strings and the Early Universe''. 
\vspace{0.5cm}


\appendix

\section{Field content of the fundamental eight dimensional massive superfield as representations of  $SO(16)$, $SO(9)$, $SO(8)$, $SO(7)$, $SO(6)$, $SU(2)^3$ and $SO(3)\otimes SO(6)$}\label{app:groupembeddingsD8}

\subsection*{The group $SO(16)$}
One can write the field content of the multiplet as representations of a group as big as $SO(16)$. The mathematical reason this embedding exists is that the massive on-shell supersymmetry algebra in eight dimensions can be mapped isomorphically to the gamma matrix algebra in $16$ dimensions. One place in the literature where this is discussed is \cite{deWit:2002vz}. An easy way to see this is to note that in both cases the representation theory reduces by the standard arguments to $8$ copies of the fermionic harmonic oscillator. Hence the massive multiplet in ten dimensions can be embedded into the chiral and anti-chiral spinor representations of $SO(16)$ ($D(8)$) with Dynkin labels
\begin{equation}
(0,\ldots,1,0)_{256} \qquad (0,\ldots,0,1)_{256}
\end{equation}
Here one spinor contains the bosons and the other the fermions. Note that the massive eight dimensional superfields used in the main text cannot be written in terms of these representations level-by-level. In the massless case the same argument as mentioned above leads to the group $SO(8)$. 

The author is not aware of any application of this in an interacting theory. In fact, since the second excited level of the open superstring is the vector of $SO(9)$ times the fundamental multiplet, the resulting list of multiplets does not embed into a sum of representations of $SO(16)$ massive case based on simple dimension counting. 

\subsection*{The group $SO(9)$}
From a physical perspective it is not surprising that an embedding of the field content of the multiplet into $SO(9)$ exists, since the minimal supersymmetries in eight and ten dimensions coincide. From \cite{greenschwarzwitten} the field content of the massive fundamental multiplet is 
\begin{equation}
(1,0,0,1)_{128} \oplus \left[ (2,0,0,0)_{44} \oplus (0,0,1,0)_{84} \right]\qquad \textrm{(SO(9))}
\end{equation} where the first factor are the fermions and the last two are the bosons. Furthermore, the first of these is the symmetric 2 index tensor, while the second is the 3 index anti-symmetric tensor. From this knowledge one can calculate the decomposition into $SO(7) \otimes SO(2)$ rather straightforwardly,
\begin{equation}\label{eq:decompso9intoso7open}\begin{array}{cl}
(2,0,0,0)_{44}   &\rightarrow  (2,0,0)^{(0)}_{27} + (1,0,0)^{\pm 2}_{7} + (0,0,0)^{(0)}_1+ (0,0,0)^{(\pm 4)}_1  \\
(0,0,1,0)_{84}   &\rightarrow  (1,0,0)^{(0)}_{7} + (0,0,2)^{(0)}_{35} + (0,1,0)^{(\pm 2)}_{21}\\
(1,0,0,1)_{128} &\rightarrow  (0,0,1)^{(\pm 1)}_8 + (1,0,1)^{(\pm 1)}_{48} + (0,0,1)^{(\pm 3)}_8 
\end{array}\end{equation}
where the superscript refer to the charge under $SO(2)$.

\subsection*{The group $SO(8)$}
There is also a direct argument for the embedding of the massive fundamental multiplet into representations of $SO(8)$: one can check from the tables in \cite{patera} that the $(0,0,0,1)$ highest weight representation of $SO(8)$ decomposes into one $(0,0,1)$ representation of $SO(7)$. Physically this is just the fact that a massless spinor in ten dimensions can be decomposed into a massive spinor in $8$ dimensions. The other levels in the multiplet are then anti-symmetrized tensor products of the  $(0,0,0,1)$ of $SO(8)$ and can in particular always be reinterpreted as representations of $SO(8)$ level-by-level. The right Dynkin labels of the multiplets can be found by advanced numerology. For this one computes the tensor product and expresses this as a sum of single representations which occur in this product whose dimensions sum to the proper binomial coefficient. The constraint to single representations arises from the fact that each $SO(7)$ representation only occurs once at each fermionic weight level in equation \eqref{eq:dynkinlabelsso7}. Having found one level, the next may then be computed by tensoring this result again with  $(0,0,0,1)$. This yields 
\begin{equation}\label{eq:so8contentopenstrchir}
\begin{array}{ccl}
0 & \quad &   (0,0,0,0)_{1}\\
1 & \quad &   (0,0,0,1)_{8} \\
2 & \quad &   (0,1,0,0)_{28} \\
3 & \quad &   (1,0,1,0)_{56} \\
4 & \quad &   (2,0,0,0)_{35}   +  (0,0,2,0)_{35} 
\end{array}
\end{equation}
as the field content in terms of $SO(8)$ representations of the fundamental massive multiplet in eight dimensions. The unwritten fermionic levels ($5-8$) repeat the pattern of the previous ones in reverse order. 

In the text the conjugate superfield based on the anti-chiral spinor representation $(0,0,1,0)$ is also needed. This is subtly different:
\begin{equation}\label{eq:so8contentopenstrantichir}
\begin{array}{ccl}
0 & \quad &   (0,0,0,0)_{1}\\
1 & \quad &   (0,0,1,0)_{8} \\
2 & \quad &   (0,1,0,0)_{28} \\
3 & \quad &   (1,0,0,1)_{56} \\
4 & \quad &   (2,0,0,0)_{35}   +  (0,0,0,2)_{35} 
\end{array}
\end{equation}
Again, the unwritten fermionic levels ($5-8$) repeat the pattern of the previous ones in reverse order. 

\subsection*{The group $SO(7)$}
This case was given already in the main text in equation \eqref{eq:dynkinlabelsso7} and is repeated here for completeness,
\begin{equation}
\begin{array}{ccl}
0 & \quad &   (0,0,0)_{1}\\
1 & \quad &   (0,0,1)_{8} \\
2 & \quad &   (0,1,0)_{21} + (1,0,0)_{7} \\
3 & \quad &   (1,0,1)_{48} + (0,0,1)_{8} \\
4 & \quad &   (2,0,0)_{27} + (0,0,2)_{35} + (1,0,0)_{7} + (0,0,0)_{1} \\ 
\end{array}
\end{equation}
where the unlisted fermionic levels repeat the previous ones in reverse order.

\subsection*{The group $SO(6)$}
The same method as that used to obtain the $SO(7)$ representation in equation \eqref{eq:dynkinlabelsso7} can be used to calculate the $SO(6)$ content of the above massive fundamental representation directly. This is relevant for comparison to light-cone methods for instance. This yields 
\begin{equation}
\begin{array}{ccl}
0 & \quad &   (0,0,0)_{1}\\
1 & \quad &   (0,0,1)_{4} + (0,1,0)_{4}\\
2 & \quad &   (0,1,1)_{15} + 2 \, (1,0,0)_{6} + (0,0,0)_{1} \\
3 & \quad &   (1,1,0)_{20} + (1,0,1)_{20} +  2 \, (0,0,1)_{4} + 2 \, (0,1,0)_{4}\\
4 & \quad &   (2,0,0)_{20} + (0,0,2)_{10} + (0,2,0)_{10} + (0,1,1)_{15} + 2\,  (1,0,0)_{6} + 3 \, (0,0,0)_{1}\\ 
\end{array}
\end{equation}
where again the unwritten fermionic levels repeat the pattern of the previous ones in reverse order.

\subsection*{The group $SO(3)\times SO(4) \sim SU(2)^3$}
The four dimensional field content of the eight dimensional multiplet can be calculated in terms of the group decomposition $SO(8) \rightarrow SO(3)\times SO(4) \sim SU(2)^3 $.  The Dynkin labels of the three $SU(2)$ groups can be calculated directly from the eight dimensional weight labels. After this the dominant weights can be sorted into highest weights as before. In a notation where the first Dynkin label corresponds to the space-time spin group and the following two to the other two $SU(2)$'s the field content can be listed as
\begin{equation}
\begin{array}{ccl}
0 & \quad &   (0,0,0)_{1}\\
1 & \quad &   (1,1,0)_{4} + (1,0,1)_{4}\\
2 & \quad &   (2,1,1)_{12} + 2 \, (2,0,0)_{3} + (0,0,2)_{3} + (0,2,0)_{3}  + (0,1,1)_{4} \\
3 & \quad &   (3,0,1)_{8} + (3,1,0)_{8} +  (1,2,1)_{12} +  (1,1,2)_{12} + 2 \, (1,0,1)_{4} + 2 (1,1,0)_{4} \\
4 & \quad &   (4,0,0)_{5} + (2,2,0)_{9} + (2,0,2)_{9} + 2 \, (2,1,1)_{12} +  (2,0,0)_3 + (0,2,2)_{9}  + 2 \, (0,1,1)_{4}   \\ & & + 3 \, (0,0,0)_{1}\\ 
\end{array}
\end{equation}
Again, the unwritten fermionic levels ($5-8$) repeat the pattern of the previous ones in reverse order. The total field content of the fundamental massive $\mathcal{N}=4$ multiplet in four dimensions is therefore
\begin{equation}
\begin{array}{cc}
\textrm{spin }0 & 42\\
\textrm{spin }\frac{1}{2}  & 48\\
\textrm{spin }1  & 123 \\
\textrm{spin }\frac{3}{2} & 8 \\
\textrm{spin } 2   & 1 
\end{array}
\end{equation}
Note this classification ignores the fact that the fields are charged under the R-symmetry group. Since this is $SU(4) \sim SO(6)$, which is related to the embedding of the eight dimensional theory in the ten dimensional one, it is instructive to work out the field content for this as well.

\subsection*{The group $SO(3)\times SO(6)$}
The $SO(4)$ content of the above can be rewritten in terms of $SO(6)$, basically because the SO(4) representations $(0,1) \oplus (0,1)$ embed in a single $(0,0,1)$ or $(0,1,0)$ spinor of $SO(6)$. Picking one fixing the rest of the multiplet as these are the anti-symmetrized tensor products of this multiplet with itself,
\begin{equation}
\begin{array}{ccl}
0 & \quad &   (0)\otimes(0,0,0)_{1}\\
1 & \quad &   (1)\otimes(0,0,1)_{8} \\
2 & \quad &   (2)\otimes(1,0,0)_{18} + (0) \otimes(0,0,2)_{10} \\
3 & \quad &   (3)\otimes(0,1,0)_{16} + (1) \otimes(1,0,1)_{40}  \\
4 & \quad &   (4)\otimes(0,0,0)_{5}  +  (2)\otimes(0,1,1)_{45} +  (0)\otimes(2,0,0)_{20}  \\ 
\end{array}
\end{equation}
These Dynkin labels were obtained by numerology\footnote{A mistake in this list in an earlier version of this paper was removed after the appearance of \cite{Feng:2012bb}} . This displays the full R-symmetry group of the massive $\mathcal{N}=4$ supersymmetric multiplet in four dimensions.


\section{Field content of the fundamental ten dimensional massive superfield as representations of $SO(32)$, $SO(16)$, $SO(10)$, $SO(9)$, $SO(8)$ and $SO(3) \otimes SO(7)$}\label{app:groupembeddingsD10}

\subsection*{The group $SO(32)$}
One can write the field content of the multiplet as representations of a group as big as $SO(32)$. The mathematical reason this embedding exists is that the massive on-shell supersymmetry algebra can be mapped isomorphically to the gamma matrix algebra in $32$ dimensions. This is discussed for instance in \cite{deWit:2002vz}. An easy way to see it is to note that in both cases the representation theory reduces by the standard arguments to $16$ copies of the fermionic harmonic oscillator. Hence the massive multiplet in ten dimensions can be embedded into the chiral and anti-chiral spinor representations of $SO(32)$ ($D(16)$) with Dynkin labels
\begin{equation}
(0,\ldots,1,0)_{32768} \qquad (0,\ldots,0,1)_{32768}
\end{equation}
Here one spinor contains the bosons and the other the fermions. Note that the massive IIB superfields used in the main text cannot be written in terms of these representations level-by-level. In the massless case the same argument as mentioned above leads to the group $SO(16)$. 

The author is not aware of any application of this in an interacting theory. In fact, since the second excited level of the open superstring is the vector of $SO(9)$ times the fundamental multiplet, the resulting list of multiplets does not embed into a sum of representations of $SO(32)$ based on simple dimension counting.

\subsection*{The group $SO(16)$}
The $\underline{16}$ of $SO(9)$ embeds into the $\underline{16}$ vector representation of $SO(16)$. Every fermionic level is now an anti-symmetrized tensor product of this $SO(16)$ representation. The explicit representations can be identified easily with the same method as above and we obtain the neat list
\begin{equation}
\begin{array}{ccl}
0 & \quad &   (0,0,0,0,0,0,0,0)_{1}\\
1 & \quad &   (1,0,0,0,0,0,0,0)_{16} \\
2 & \quad &   (0,1,0,0,0,0,0,0)_{120} \\
3 & \quad &   (0,0,1,0,0,0,0,0)_{560} \\
4 & \quad &   (0,0,0,1,0,0,0,0)_{1820}  \\
5 & \quad &   (0,0,0,0,1,0,0,0)_{4368}  \\
6 & \quad &   (0,0,0,0,0,1,0,0)_{8008}  \\
7 & \quad &   (0,0,0,0,0,0,1,1)_{11440} \\
8 & \quad &   (0,0,0,0,0,0,2,0)_{6435} +   (0,0,0,0,0,0,0,2)_{6435}  \\
\end{array}
\end{equation}
for the first $9$ levels of the superfield, suppressing the levels $9$ to $16$ which as above follow this pattern backwards. The total field content of the massive fundamental multiplet can be written as the tensor products
\begin{equation}
\textrm{total} = \left[ (0,0,0,0,0,0,0,1)_{128} + (0,0,0,0,0,0,1,0)_{128}\right]^2
\end{equation}
where the bosonic and fermionic fields can be isolated as
\begin{eqnarray}
\textrm{bosonic} &= & \left[(0,0,0,0,0,0,0,1)_{128}\right]^2 +  \left[(0,0,0,0,0,0,1,0)_{128}\right]^2 \\ 
\textrm{fermionic} &= & 2 \left[  (0,0,0,0,0,0,0,1)_{128} \otimes (0,0,0,0,0,0,1,0)_{128} \right]
\end{eqnarray}
Moreover, one can embed the massless fields of IIB into $2$ representations of $SO(16)$ as well which works partly since for instance
\begin{equation}
\begin{array}{ccc}
(0,0,0,0,0,0,0,1)_{D8} & = & (0,0,0,1)_{D4} \otimes (0,0,0,1)_{D4} + (0,0,1,0)_{D4} \otimes (0,0,1,0)_{D4} \\
(0,0,0,0,0,0,1,0)_{D8} & = & 2 \,\, (0,0,0,1)_{D4} \otimes (0,0,1,0)_{D4}
\end{array}
\end{equation}
which was obtained from \cite{patera} for a standard embedding. If we now take the quantum numbers for the diagonal sum of the two $D4$ algebras we obtain
\begin{equation}
\begin{array}{ccc}
(0,0,0,1)\otimes (0,0,0,1) & = &  (0,0,0,2)_{35} + (0,1,0,0)_{28} +  (0,0,0,0)_{1} \\
(0,0,1,0)\otimes (0,0,1,0) & = & (0,0,2,0)_{35} + (0,1,0,0)_{28} +  (0,0,0,0)_{1} \\
(0,0,0,1)\otimes (0,0,1,0) & = & (0,0,1,1)_{56} + (0,0,0,1)_{8}
\end{array}
\end{equation}
using \cite{lie}. Comparing to \eqref{eq:listofmasslessSO8reps} it is seen that although the dimensions nicely match up the Dynkin labels do not quite. These would match completely if the tables in \cite{patera} would have $(0,0,0,1)$ replaced by  $(1,0,0,0)$. Since the triality group acts as interchanges of the 1st,3rd and 4th Dynkin labels, a non-standard embedding of the two $SO(8)$ groups into $SO(16)$ will work.

\subsection*{The group $SO(10)$}
There is also a direct argument for the embedding of the massive fundamental multiplet into representations of $SO(10)$: one can check from the tables in \cite{patera} that the $(0,0,0,1,0)$ highest weight representation of $SO(10)$ decomposes into one $(0,0,0,1)$ representation of $SO(9)$. The other levels in the multiplet are then anti-symmetrized tensor products of  $(0,0,0,1,0)$ and can hence always be reinterpreted as representations of $SO(10)$. The right Dynkin labels of the multiplets can be found by advanced numerology. For this one computes the tensor product and expresses this as a sum of single representations which occur in this product whose dimensions sum to the proper binomial coefficient. The constraint to single representations arises from the fact that each $SO(9)$ representation only occurs once at each fermionic weight level in equation \eqref{eq:dynkinlabelsso9}. Having found one level, the next may then be computed by tensoring this result again with  $(0,0,0,1,0)$. This yields 
\begin{equation}
\begin{array}{ccl}
0 & \quad &   (0,0,0,0,0)_{1}\\
1 & \quad &   (0,0,0,1,0)_{16} \\
2 & \quad &   (0,0,1,0,0)_{120} \\
3 & \quad &   (0,1,0,0,1)_{560} \\
4 & \quad &   (0,2,0,0,0)_{770}   +  (1,0,0,0,2)_{1050}\end{array}\qquad \begin{array}{ccl}
5 & \quad &   (0,0,0,0,3)_{672}   + (1,1,0,0,1)_{3696} \\
6 & \quad &   (0,1,0,0,2)_{3696} + (2,0,1,0,0)_{4312} \\
7 & \quad &   (3,0,0,1,0)_{2640} + (1,0,1,0,1)_{8800} \\
8 & \quad &   (0,1,0,1,1)_{5940} + (0,0,1,0,2)_{6930} 
\end{array}
\end{equation}
as the field content in terms of $SO(10)$ representations of the fundamental massive multiplet in IIB/A. It can be checked by dimension numerology for instance there is no combination of representations of $E_6$ which would the above list of $SO(10)$ representations. Also, with the help of the same numerology and \cite{lie} there is no tensor product of two sums of $SO(10)$ reps which yields the above list of representations.

The outlined procedure turns out to be ambiguous at the final step as it yields two possible sums. One of these involves the representation $(4,0,0,0,0)_{660}$, which decomposes (\cite{patera})  into $SO(9)$ representations which do not all occur in the result found above. Hence this possibility can be discarded, yielding the table above. The further levels ($9-16$) repeat the found levels in reverse order. Note that every representation occurs only once in the above list. Including fermionic weight levels not displayed this shows there are only two scalars in the multiplet at the top and bottom state.

\subsection*{The group $SO(9)$}
This case was given already in the main text in equation \eqref{eq:dynkinlabelsso9} and is repeated here for completeness,
\begin{equation}
\begin{array}{ccl}
0 & \quad &   (0,0,0,0)_{1}\\
1 & \quad &   (0,0,0,1)_{16} \\ 
2 & \quad &   (0,1,0,0)_{36} + (0,0,1,0)_{84}\\
3 & \quad &   (1,0,0,1)_{128} + (0,1,0,1)_{432}\\
4 & \quad &   (2,0,0,0)_{44} + (0,0,0,2)_{126} + (1,1,0,0)_{231} + (0,2,0,0)_{495} + (1,0,0,2)_{924}\\
5 & \quad &   (1,0,0,1)_{128} + (0,1,0,1)_{432} + (2,0,0,1)_{576} + (0,0,0,3)_{672}+ (1,1,0,1)_{2560} \\
6 & \quad &   (0,1,0,0)_{36} + (0,0,1,0)_{84}   + (1,1,0,0)_{231} + (1,0,1,0)_{594} + (1,0,0,2)_{924}  \\ & & + (2,1,0,0)_{910} + (2,0,1,0)_{2457} + (0,1,0,2)_{2772} \\
7 & \quad &   (0,0,0,1)_{16} + (1,0,0,1)_{128} + (0,1,0,1)_{432} + (2,0,0,1)_{576} \\ & & + (0,0,1,1)_{768} + (3,0,0,1)_{1920}+ (1,1,0,1)_{2560} + (1,0,1,1)_{5040}\\
8 & \quad &   (0,0,0,0)_{1} + (1,0,0,0)_{9} + (0,0,1,0)_{84} + (2,0,0,0)_{44} + (0,0,0,2)_{126} \\ & & + (1,0,1,0)_{594} + (0,2,0,0)_{495}   + (1,0,0,2)_{924} + (3,0,0,0)_{156} +  (0,1,1,0)_{1650} \\ & & + (2,0,1,0)_{2457}  + (2,0,0,2)_{3900}  + (0,0,2,0)_{1980} + (4,0,0,0)_{450}
\end{array}
\end{equation}
where the unlisted fermionic levels repeat the previous ones in reverse order.

\subsection*{The group $SO(8)$}\label{subsec:so8closed}
The field content of the superfield can also be calculated in terms of $SO(8)$ directly which can be useful  for comparison to light-cone methods. This yields for the bosonic part of the multiplet
\begin{equation}
\begin{array}{ccl}
0 & \quad &   (0,0,0,0)_{1}\\
2 & \quad &   (1,0,0,0)_{8} + 2 \, (0,1,0,0)_{28} + (0,0,1,1)_{56} \\
4 & \quad &   (0,0,0,0)_{1} + 2 \, (1,0,0,0)_{8} + (0,1,0,0)_{28}  + 3 \, (2,0,0,0)_{35}   + 2 \, (0,0,2,0)_{35} \\ & &+ 2 \, (0,0,0,2)_{35}  + 2 \, (0,0,1,1)_{56} + 2 \, (1,1,0,0)_{160}   +  (1,0,2,0)_{224} + (1,0,0,2)_{224} \\ & &  +  (0,2,0,0)_{300}+ (1,0,1,1)_{350} \\
6 & \quad &   3 \, (1,0,0,0)_{8} + 6 \, (0,1,0,0)_{28}  + 2 \, (2,0,0,0)_{35}    +  (0,0,2,0)_{35} +  (0,0,0,2)_{35} + 4 \, (0,0,1,1)_{56}  \\ & &+  4\, (1,1,0,0)_{160}   +  2 \, (1,0,2,0)_{224}  + 2 \,(1,0,0,2)_{224}  + 4 \, (1,0,1,1)_{350} + (3,0,0,0)_{112} \\ & &   + (0,1,1,1)_{840}+ 2 \, (2,1,0,0)_{567} + (0,1,2,0)_{567} + (2,0,1,1)_{1296} + (0,1,0,2)_{567}   \\
8 & \quad &    5 \, (0,0,0,0)_1 +  4 \, (1,0,0,0)_{8} + 3 \, (0,1,0,0)_{28}  + 4 \, (2,0,0,0)_{35}    + 3 \,  (0,0,2,0)_{35}  \\ & &+ 3 \,  (0,0,0,2)_{35} + 6 \, (0,0,1,1)_{56} +  4\, (1,1,0,0)_{160}    +  2 \, (1,0,2,0)_{224} + 2 \, (1,0,0,2)_{224}   \\ & & + 5 \, (1,0,1,1)_{350} + 3 \, (0,2,0,0)_{300}+  2 \, (3,0,0,0)_{112} + 2 \, (0,1,1,1)_{840}   + (2,1,0,0)_{567}    \\ & &+ 2 \, (2,0,1,1)_{1296}   \end{array}
\end{equation}
and for the fermionic part
\begin{equation}
\begin{array}{ccl}
1 & \quad &   (0,0,0,1)_{8} +  (0,0,1,0)_8 \\
3 & \quad &   \left[(0,1,0,1)_{160} + 2\, (1,0,1,0)_{56} + (0,0,0,1)_8 \right] + \left[(0,1,1,0)_{160} + 2\, (1,0,0,1)_{56} + (0,0,1,0)_8 \right]\\
5 & \quad &   \left[(0,0,0,3)_{112} + (1,1,1,0)_{840} + (0,0,2,1)_{224} + 2\, (2,0,0,1)_{224} + 4\,(1,0,1,0)_{56} \right. \\ & & \left. + 2\,(0,0,0,1)_8   \right] + \left[(0,0,3,0)_{112} + (1,1,0,1)_{840} + (0,0,1,2)_{224} + 2\, (2,0,1,0)_{224}  \right. \\ & & \left. + 4\,(1,0,0,1)_{56}+ 4\,(0,0,1,0)_8   \right]  \\
7 & \quad &  \left[ (1,0,1,2)_{1296} + (3,0,1,0)_{672} + 2\, (1,1,1,0)_{840} + 2\, (0,0,2,1)_{224} +3\, (2,0,0,1)_{224}  \right. \\ & & \left.  + 5 \, (1,0,1,0)_{56} + 4\, (0,0,0,1)_8 \right] +  \left[ (1,0,2,1)_{1296} + (3,0,0,1)_{672} + 2\, (1,1,0,1)_{840} \right. \\ & & \left.+ 2\, (0,0,1,2)_{224} +3\, (2,0,1,0)_{224}   + 5\, (1,0,0,1)_{56} + 4\, (0,0,1,0)_8 \right]
\end{array}
\end{equation}
where it is emphasized the fermions can be interpreted as the decomposition of an $SO(8)$ Dirac-type spinor into it's Weyl components: the multiplets are symmetric under interchange of the last two Dynkin labels. 

The $SO(8)$ field content of the fundamental superfield can also be written as a product of $SO(8)$ representations level by level. In the main text this is discussed as an example of the KLT relations for free fields expressed in superfields. 
\begin{equation}\label{eq:KLTforfreefieldfieldcontent}
\begin{array}{ccl}
0 & \quad &   (0,0,0,0)_{1} \otimes  (0,0,0,0)_{1} \\
1 & \quad &   (0,0,0,1)_{8} \otimes  (0,0,0,0)_{1}  + (0,0,0,0)_{1} \otimes  (0,0,1,0)_{8} \\
2 & \quad &   (0,1,0,0)_{28} \otimes  (0,0,0,0)_{1}  + (0,0,0,1)_{8} \otimes  (0,0,1,0)_{8}  + (0,0,0,0)_{1} \otimes  (0,1,0,0)_{28}  \\
3 & \quad &   (1,0,1,0)_{56} \otimes  (0,0,0,0)_{1}  + (0,1,0,0)_{28} \otimes  (0,0,1,0)_{8}  + (0,0,0,1)_{8} \otimes  (0,1,0,0)_{28} + \\ & & (0,0,0,0)_{1} \otimes  (1,0,0,1)_{56}  \\
4 & \quad &   \left((2,0,0,0)_{35} + (0,0,2,0)_{35} \right) \otimes (0,0,0,0)_{1}  + (1,0,1,0)_{56} \otimes (0,0,1,0)_{8}  + (0,1,0,0)_{28} \otimes   \\ & & (0,1,0,0)_{28} +  (0,0,0,1)_{8} \otimes (1,0,0,1)_{56} + (0,0,0,0)_{1} \otimes   \left((2,0,0,0)_{35} + (0,0,0,2)_{35} \right)  \\
5 & \quad &   (1,0,1,0)_{56} \otimes  (0,0,0,0)_{1}  +   \left((2,0,0,0)_{35} + (0,0,2,0)_{35} \right) \otimes (0,0,1,0)_{8} + \\ & &  (1,0,1,0)_{56} \otimes  (0,1,0,0)_{28}  + (0,1,0,0)_{28} \otimes  (1,0,0,1)_{56} +  \\ & &  (0,0,0,1)_{8}  \otimes   \left((2,0,0,0)_{35} +(0,0,0,2)_{35} \right) (0,0,0,0)_{1} \otimes  (1,0,0,1)_{56}  \\
6 & \quad &  (0,1,0,0)_{28} \otimes  (0,0,0,0)_{1} + (1,0,1,0)_{56} \otimes (0,0,1,0)_{8} + \left((2,0,0,0)_{35} + (0,0,2,0)_{35} \right) \otimes  \\ & & (0,1,0,0)_{28}    + (1,0,1,0)_{56} \otimes  (1,0,0,1)_{56}  +  (0,1,0,0)_{28} \otimes   \left((2,0,0,0)_{35} + (0,0,0,2)_{35} \right)  \\ & & +  (0,0,0,1)_{8} \otimes (1,0,0,1)_{56}  + (0,0,0,0)_{1} \otimes  (0,1,0,0)_{28} \\
7 & \quad &   (0,0,0,1)_{8} \otimes  (0,0,0,0)_{1}  +  (0,1,0,0)_{28} \otimes  (0,0,1,0)_{8}  +  (1,0,1,0)_{56} \otimes  (0,1,0,0)_{28}  + \\ & &  \left((2,0,0,0)_{35} + (0,0,2,0)_{35} \right) \otimes (1,0,0,1)_{56} + (1,0,0,1)_{56}  \otimes \left((2,0,0,0)_{35} + (0,0,0,2)_{35} \right) +  \\ & &   (0,1,0,0)_{28} \otimes  (1,0,0,1)_{56} +  (0,0,0,1)_{8} \otimes  (0,1,0,0)_{28} + (0,0,0,0)_{1} \otimes  (0,0,1,0)_{8}\\
8 & \quad &   2  \, \left[ (0,0,0,0)_{1} \otimes  (0,0,0,0)_{1} +  (0,0,0,1)_{8} \otimes  (0,0,1,0)_{8} + (0,1,0,0)_{28} \otimes  (0,1,0,0)_{28} +\right. \\ & & \left. (1,0,1,0)_{56} \otimes  (1,0,0,1)_{56} \right] + \left((2,0,0,0)_{35} + (0,0,2,0)_{35} \right) \otimes \left((2,0,0,0)_{35} + (0,0,0,2)_{35} \right) 
 \end{array}
\end{equation}

\subsection*{The group $SO(3)\otimes SO(7)$}
For decomposition to lower dimensions one might be interested in decomposing the above into a group with a $SO(3)$ factor. It is useful to use a large second group to keep the representations manageable.  Here we opt to use $SO(7)$. The bosonic levels read
\begin{equation}
\begin{array}{ccl}
0 & \quad &   (0)_1 \otimes(0,0,0)_{1}\\
2 & \quad &   (2)_3 \otimes [ (0,1,0)_{21} +  (1,0,0)_{7}  ] + (0)_1 \otimes [ (0,0,2)_{35} + (0,0,0)_{1}   ]  \\
4 & \quad &   (4)_5 \otimes [(2,0,0)_{27} + (0,0,2)_{35} +  (1,0,0)_{7} + (0,0,0)_{1} ] + (2)_3 \otimes [  (1,0,2)_{189} + (1,1,0)_{105}  \\ & & + (0,0,2)_{35} + 2 \,  (0,1,0)_{21} + (1,0,0)_{7}  ]  + (0)_1 \otimes [  (0,2,0)_{168}   + (1,1,0)_{105} + (2,0,0)_{27} \\ & & +  (0,0,2)_{35} + (0,0,0)_{1}   ]   \\ 
6 & \quad & (6)_7 \otimes \textbf{\large [} (0,1,0)_{21} +(1,0,0)_{7} ] + (4)_5 \otimes [ (1,0,2)_{189}  +  (1,1,0)_{105} + (2,0,0)_{27} \\ & & + 2 \, (0,0,2)_{35}  +  (0,1,0)_{27} + (1,0,0)_{7} + (0,0,0)_{1} ] + (2)_3 \otimes [ (0,1,2)_{378} +  (2,1,0)_{330}  \\ & & + (3,0,0)_{77}  + 2 \, (1,0,2)_{189}  + 2 \, (1,1,0)_{105} +  (2,0,0)_{27} + (0,0,2)_{35} + 3 \,  (0,1,0)_{21}  \\ & & + 2 \, (1,0,0)_{7}  ] +  (0)_1 \otimes [  (2,0,2)_{616} + (0,2,0)_{168} + (1,0,2)_{189} + (1,1,0)_{105}+ (2,0,0)_{27}  \\ & &  + 2 \, (0,0,2)_{35} + (0,0,0)_{1}   ]
 \\
8 & \quad & (8)_{9} \otimes \left[(0,0,0)_1\right] + (6)_7 \otimes \textbf{\large [} (0,0,2)_{35} +(0,1,0)_{21} +(1,0,0)_{7}  ] + (4)_5 \otimes \left[(0,2,0)_{168}  \right.  \\ & &  \left. +(1,0,2)_{189} + 2 \, (1,1,0)_{105} + 2 \, (2,0,0)_{27}+ 2 \, (0,0,2)_{35} +   (0,1,0)_{21} +(1,0,0)_{7} \right. \\ & & \left. + (0,0,0)_{1} \right]  + (2)_3 \otimes [  (2,0,2)_{616} + (0,1,2)_{378} +  (2,1,0)_{330} +  (3,0,0)_{77}  +3 \, (1,0,2)_{189}  \\ & & + 3 \, (1,1,0)_{105}  + (2,0,0)_{27}+ 2 \, (0,0,2)_{35} + 3 \,  (0,1,0)_{21} + 2 \, (1,0,0)_{7} ]  \\  & & + (0)_1 \otimes [ (4,0,0)_{182} + (0,0,4)_{294}  + (2,0,2)_{616} + (3,0,0)_{77} + (0,2,0)_{168}  + (1,0,2)_{189} + \\ & &(1,1,0)_{105}  + 2 \, (2,0,0)_{27}+ 2 \, (0,0,2)_{35} + (1,0,0)_{7}  + 2 \, (0,0,0)_{1}   ]
\end{array}
\end{equation}
while the field content on the fermionic levels reads
\begin{equation}
\begin{array}{ccl}
1 & \quad &   (1)_2 \otimes (0,0,1)_{8} \\
3 & \quad &   (3)_4 \otimes  [ (1,0,1)_{48}  + (0,0,1)_8]  +  (1)_2 \otimes [  (0,1,1)_{112} + (1,0,1)_{48} + (0,0,1)_8 ] \\
5 & \quad &   (5)_6 \otimes [(1,0,1)_{48} + (0,0,1)_8] +  (3)_4 \otimes  [ (0,0,3)_{112} +  (2,0,1)_{168} +  (0,1,1)_{112}  \\ & & + 2 \,  (1,0,1)_{48} ] +  (1)_2 \otimes [ (1,1,1)_{512} + (2,0,1)_{168}  + 2 \, (0,1,1)_{112} + 2 \, (1,0,1)_{48}  \\ & & + 3 \, (0,0,1)_8 ] \\
7 & \quad &   (7)_8 \otimes [(0,0,1)_8] + (5)_6 \otimes [(0,1,1)_{112} + 2 \, (1,0,1)_{48} + (0,0,1)_8] +  (3)_4 \otimes [(1,1,1)_{512} \\ & & + (0,0,3)_{112} + 2 \, (2,0,1)_{168} + 2 \, (0,1,1)_{112} + 3 \,  (1,0,1)_{48} + 4 \, (0,0,1)_8] \\ & & +  (1)_2 \otimes [(1,0,3)_{560} + (3,0,1)_{448} + (1,1,1)_{512} + (0,0,3)_{112} + 2 \, (2,0,1)_{168} \\ & & + 2 \, (0,1,1)_{112} + 3 \, (1,0,1)_{48} ] 
\end{array}
\end{equation}
Here again the remaining levels repeat the same pattern in reverse. The full multiplet contains $4274$ scalars (disregarding R-symmetry representations), and one spin $4$ field. The $SO(7)$ field content can be embedded into an $SO(8)$ since there is an embedding for the first level
\begin{equation}
(0,0,1) \rightarrow (0,0,0,1)
\end{equation}
and the rest of the multiplet is the anti-symmetrized tensor product of this state. Similarly, the first level can be rewritten as
\begin{equation}
(1)_2 \otimes (0,0,1)_{8} \rightarrow (2)_3 \otimes (1,0,0,0,0,0)_{13}
\end{equation}
in terms of an $SO(3)\otimes SO(13)$, guaranteeing that the full multiplet can be embedded into this group level by level.

\bibliographystyle{JHEP}

\bibliography{10Dbibmass}
\end{document}